\pdfoutput=1 
\documentclass[12pt,aps,prl,reprint,superscriptaddress,amsmath,amsfonts,amssymb]{revtex4-1}
\usepackage[]{revquantum}
\usepackage[english]{babel}
\usepackage{mathrsfs}
\usepackage{hyperref}
\usepackage{hypernat}
\usepackage[utf8]{inputenc}
\usepackage[caption=false]{subfig}
\usepackage{graphicx}
\graphicspath{{./Figures/}}
\begin{document}
\title{Out-of-Time-Ordered-Correlator Quasiprobabilities Robustly Witness Scrambling}
\author{José Raúl González Alonso}
\email[Corresponding author: ]{gonzalezalonso@chapman.edu}
\affiliation{Schmid College of Science and Technology, Chapman University, Orange, California 92866, USA}
\author{Nicole Yunger Halpern}
\affiliation{Institute for Quantum Information and Matter, Caltech, Pasadena, California 91125, USA}
\author{Justin Dressel}
\affiliation{Schmid College of Science and Technology, Chapman University, Orange, California 92866, USA}
\affiliation{Institute for Quantum Studies, Chapman University, Orange, California 92866, USA}

\date{\today}

\begin{abstract}
 Out-of-time-ordered correlators (OTOCs) have received considerable recent attention as qualitative witnesses of information scrambling in many-body quantum systems.
 Theoretical discussions of OTOCs typically focus on closed systems, raising the question of their suitability as scrambling witnesses in realistic open systems.
 We demonstrate empirically that the nonclassical negativity of the quasiprobability distribution (QPD) behind the OTOC is a more sensitive witness for scrambling than the OTOC itself.
 Nonclassical features of the QPD evolve with timescales that are robust with respect to decoherence and are immune to false positives caused by decoherence.
 To reach this conclusion, we numerically simulate spin-chain dynamics and three measurement protocols (the interferometric, quantum-clock, and weak-measurement schemes) for measuring OTOCs. We target experiments based on quantum-computing hardware such as superconducting qubits and trapped ions.
\end{abstract}

\maketitle

\emph{Introduction}---Quantum many-body dynamics is \emph{scrambling} when initially localized quantum information spreads via entanglement through many degrees of freedom. Out-of-time-ordered correlators (OTOCs) have been suggested as a way to characterize scrambling across
condensed-matter and high-energy contexts \cite{Larkin_QuasiclassicalMethodTheory_1969,Kitaev_HiddencorrelationsHawking_2014,Shenker_Blackholesbutterfly_2014,Shenker_Multipleshocks_2014,Hartnoll_Theoryuniversalincoherent_2015,Shenker_Stringyeffectsscrambling_2015,Kitaev_SimpleModelQuantum_2015,Roberts_Localizedshocks_2015,Roberts_DiagnosingChaosUsing_2015,Maldacena_boundchaos_2016,Aleiner_Microscopicmodelquantum_2016,Blake_UniversalChargeDiffusion_2016,Blake_Universaldiffusionincoherent_2016,Chen_UniversalLogarithmicScrambling_2016,Lucas_Chargediffusionbutterfly_2016,Roberts_LiebRobinsonBoundButterfly_2016,Hosur_Chaosquantumchannels_2016,Banerjee_Solvablemodeldynamical_2017,Fan_Outoftimeordercorrelationmanybody_2017,Gu_Localcriticalitydiffusion_2017,Roberts_Chaoscomplexitydesign_2017,Chen_Operatorscramblingquantum_2018,Huang_Outoftimeorderedcorrelatorsmanybody_2017,Iyoda_ScramblingQuantumInformation_2018,Yoshida_EfficientdecodingHaydenPreskill_2017,Lin_Outoftimeorderedcorrelatorsquantum_2018,Pappalardi_Scramblingentanglementspreading_2018,YungerHalpern_Reconcilingtwonotions_2018}.
Hence, investigating how to measure OTOCs experimentally is crucial.
Different OTOC-measurement protocols have been proposed 
\cite{Yao_InterferometricApproachProbing_2016,Swingle_Measuringscramblingquantum_2016,
YungerHalpern_QuasiprobabilityOutofTimeOrderedCorrelator_2018,Zhu_Measurementmanybodychaos_2016},
and some experimental success has been reported
\cite{Wei_ExploringLocalizationNuclear_2016,Garttner_Measuringoutoftimeordercorrelations_2017,Li_MeasuringOutofTimeOrderCorrelators_2017,Landsman_VerifiedQuantumInformation_2018}.
Yet the protocols' robustness in realistic, decoherent experimental settings
has just started to be explored and is emerging as an active area of research \cite{Syzranov_Outoftimeordercorrelatorsfinite_2017,Zhang_Informationscramblingchaotic_2018,Swingle_Resiliencescramblingmeasurements_2018,Yoshida_DisentanglingScramblingDecoherence_2018,Knap_Entanglementproductioninformation_2018,Landsman_VerifiedQuantumInformation_2018}.

We study decoherence's effects on OTOCs used to witness information scrambling. We find that the OTOCs' underlying 
quasiprobability distributions (QPDs) can more robustly identify
the key timescales that distinguish scrambling. These QPDs are extended Kirkwood-Dirac QPDs \cite{Kirkwood_QuantumStatisticsAlmost_1933,Dirac_AnalogyClassicalQuantum_1945,Terletsky_limitingtransitionquantum_1937,Margenau_CorrelationMeasurementsQuantum_1961,Chaturvedi_WignerWeylcorrespondence_2006,YungerHalpern_Jarzynskilikeequalityoutoftimeordered_2017,YungerHalpern_QuasiprobabilityOutofTimeOrderedCorrelator_2018}. They reduce to classical joint probability distributions over the 
eigenvalues of the OTOC operators when the operators commute. Otherwise, the QPDs become nonclassical: individual quasiprobabilities can become negative, exceed 
one, or become nonreal. This nonclassicality robustly distinguishes scrambling from decoherence.

We study three OTOC-measurement protocols: the
(1) interferometric \cite{Swingle_Measuringscramblingquantum_2016}, (2) sequential-weak-measurement
 \cite{YungerHalpern_Jarzynskilikeequalityoutoftimeordered_2017,YungerHalpern_QuasiprobabilityOutofTimeOrderedCorrelator_2018},
and (3) quantum-clock \cite{Zhu_Measurementmanybodychaos_2016} protocols. Scrambling causes the OTOC to decay over a short time interval, then 
remain small.
Information leakage can reproduce this behavior \cite{Zhang_Informationscramblingchaotic_2018}, since a 
decohered system entangles with the environment. Quantum information spreads across many degrees of freedom, 
but most are outside the system. We therefore propose a modification to these protocols that uses the (coarse-grained \cite{YungerHalpern_QuasiprobabilityOutofTimeOrderedCorrelator_2018}) QPD
behind the OTOC to distinguish between scrambling and
nonscrambling dynamics despite decoherence. 

Our Letter is organized as follows. We first define the OTOC and its QPD.
As a concrete example suitable for simulation with qubit architectures, we consider a spin chain switchable between scrambling and integrable dynamics. Next, we introduce dephasing, modeled on current
superconducting-qubit technology, and we analyze its effect on the OTOC and its QPD.
We numerically simulate the spin chain for each OTOC-measurement protocol, and we compare the OTOC's degradation
by decoherence. The simulations show that the QPD's negativity distinguishes scrambling
dynamics despite ambiguity in the OTOC.

\emph{OTOCs and their quasiprobabilities}---Quantum information scrambling is related to the quantum butterfly effect: localized operators'
supports grow under time evolution by an appropriate nonintegrable Hamiltonian. The operators 
come to have large commutators with most other operators---even operators localized far from the initially considered local operator. As an example, consider a Pauli operator acting on one end of a spin chain. Another Pauli operator, acting on the opposite end, probes the propagation of quantum information.
If the Hamiltonian is scrambling, an increasing number of degrees of freedom must be measured to recover the initially
local information. Below, we make this intuition and its relation to the OTOC more precise.

Let $H$ denote a quantum many-body system Hamiltonian; $W$ and $V,$ local far-apart operators;
and $\rho$, a density matrix. The OTOC is defined as
\begin{equation}\label{eq:otoc}
F(t) := \Tr\left( W^\dagger(t) V^\dagger W(t) V \rho \right).
\end{equation}
Here, $W(t) = U(t)^\dagger W U(t)$ is evolved in the Heisenberg picture with the unitary evolution operator
$U(t):=\exp(-\ii H t)$. Initially, $W$ and $V$ commute: $[W(0),V]=0$.
If $W$ and $V$ are unitary, then the OTOC is related to the Hermitian square of their commutator:
\begin{equation}
C(t) := \left\langle\frac{\left[W(t), V\right]^\dagger}{(2i^*)}\frac{\left[W(t), V\right]}{2i}
\right\rangle = \frac{1-\mathrm{Re}\;F(t)}{2}.
\end{equation}
Otherwise, the commutator's square includes nonconstant time-ordered correlators.
A Hamiltonian that scrambles information tends to grow the commutator's magnitude.
This growth leads to a persistent smallness of $\mathrm{Re}\; F(t)$.
In contrast, for a nonscrambling Hamiltonian, $W(t)$ and $V$ approximately commute 
after a short recurrence time, as information quickly recollects from other parts of the system.
$\mathrm{Re}\; F(t)$ revives to close to one.

$W$ and $V$ decompose as $W = \sum_{w} w \Pi_{w}^W$ and $V = \sum_{v} v \Pi_{v}^V$,
where $\Pi_{w}^W$ and $\Pi_{v}^V$ are the projectors onto the eigenspaces corresponding to the eigenvalues
$w$ and $v$. The eigenspaces are degenerate, since $W$ and $V$ are local operators
and the system is large. $F(t)$ can be expressed as an average of eigenvalues
\footnote{We index the $W$ and $V$ eigenvalues in Eq.~\eqref{eq:otoc_and_cg_qp} following the conventions in \cite{YungerHalpern_Jarzynskilikeequalityoutoftimeordered_2017,YungerHalpern_QuasiprobabilityOutofTimeOrderedCorrelator_2018}.},
\begin{equation}\label{eq:otoc_and_cg_qp}
    F(t) = \sum_{v_1, w_2, v_2, w_3} v_1 w_2 v_2^* w_3^* \, \tilde{p}_t \left(v_1, w_2, v_2, w_3 \right),
\end{equation}
with respect to an extended Kirkwood-Dirac \cite{Kirkwood_QuantumStatisticsAlmost_1933,Dirac_AnalogyClassicalQuantum_1945}
\emph{(coarse-grained) quasiprobability distribution (QPD)} 
\begin{equation}\label{eq:coarse_quasi_prob}
    \tilde{p}_t \left(v_1, w_2, v_2, w_3 \right) 
        := \Tr\left( \Pi_{w_3}^{W(t)} \Pi_{v_2}^{V} \Pi_{w_2}^{W(t)} \Pi_{v_1}^{V} \rho \right).
\end{equation}
$\tilde{p}_t$ was denoted by $\tilde{\mathscr{A}}_\rho$ in
\cite{YungerHalpern_QuasiprobabilityOutofTimeOrderedCorrelator_2018}.

Equation \eqref{eq:otoc_and_cg_qp} implies that the QPD 
$\tilde{p}_t$ exhibits the OTOC's timescales. Therefore,
qualitative features of OTOCs that reflect scrambling should
have counterparts in $\tilde{p}_t$.

The QPD $\tilde{p}_t$ is complex and, like a classical probability distribution,
normalized: $\sum_{{v_1},{w_2},{v_2},{w_3}} \tilde{p}_t \left(v_1, w_2, v_2, w_3 \right)=1$.
Regions where $\tilde{p}_t$ becomes negative, exceeds one, or has a nonzero imaginary
part are nonclassical. We quantify these regions' magnitudes with the
\emph{total nonclassicality}
of $\tilde{p}_t$:
\begin{equation}
\label{eq:total_non-class}
\tilde{N}(t) := \sum_{{v_1},{w_2},{v_2},{w_3}} \left|\tilde{p}_t \left(v_1, w_2, v_2, w_3 \right)\right| - 1.
\end{equation}
As we will see, even in the presence of decoherence, the total nonclassicality's evolution
distinguishes integrable from nonintegrable Hamiltonians.
The distinction allows the QPD to signal scrambling robustly.

\emph{Spin chain}---We illustrate with a quantum Ising chain of $N$ qubits.
For ease of comparison, we use the conventions in  \cite{Berman_Delocalizationborderonset_2001,Banuls_StrongWeakThermalization_2011,Gubin_Quantumchaosintroduction_2012,Kim_BallisticSpreadingEntanglement_2013}:
\begin{equation}
H = -J \sum_{i=1}^{N-1} \sigma_{i}^{z}\sigma_{i+1}^{z} - h\sum_{i=1}^{N}\sigma_{i}^{z} - g\sum_{i=1}^{N}\sigma_{i}^{x}.
\end{equation}
We set $\hbar = 1$, such that energies are measured in units of $J$; and times, in units of $1/J$.
We fix $2\pi / J = 1\,\mu s$ and simulate two cases: (1) Integrable case:  $h/J=0.0,\, g/J=1.05$, and (2) nonintegrable case: $h/J=0.5,\, g/J=1.05$.
These values equal those in Ref.~\cite{YungerHalpern_QuasiprobabilityOutofTimeOrderedCorrelator_2018}.
As in Ref.~\cite{YungerHalpern_QuasiprobabilityOutofTimeOrderedCorrelator_2018},
$W=\sigma_1^z$, and $V=\sigma_N^z$
\footnote{Note that $\sigma_x$ with an integrable Hamiltonian can simulate scrambling \cite{Lin_Outoftimeorderedcorrelatorsquantum_2018}.}.

To map this Hamiltonian onto a physical qubit system, e.g., an array of transmons \cite{Barends_CoherentJosephsonQubit_2013,Koch_Chargeinsensitivequbitdesign_2007}, we interpret the eigenstates of
$-\sigma_i^x$ as a qubit's energy eigenbasis. Each qubit has an intrinsic energy splitting of $2g$
and couples capacitively to its neighbors with energy $J$. Unless prepared by a measurement, the qubit relaxes to a
thermal state. Therefore, as an initial state, we consider a Gibbs state at 
a finite temperature $T$: $\rho_T = \mathcal{Z}^{-1}\exp(-H/T)$, with $T/J=1$, $\mathcal{Z} = \mathrm{Tr}(\exp(-H/T))$,
and $k_B=1$ \footnote{Additionally, this allows us to circumvent the
difficulties associated with experimentally preparing an infinite-temperature Gibbs state.}.
Each qubit has a ground-state population of $\approx 0.8$.
OTOCs are usually evaluated on thermal states due to holographic interest in the thermofield double state
\cite{Shenker_Blackholesbutterfly_2014,Shenker_Multipleshocks_2014,Shenker_Stringyeffectsscrambling_2015,Kitaev_SimpleModelQuantum_2015,Roberts_Localizedshocks_2015,Roberts_DiagnosingChaosUsing_2015,Maldacena_boundchaos_2016,Blake_UniversalChargeDiffusion_2016,Blake_Universaldiffusionincoherent_2016,Hosur_Chaosquantumchannels_2016}.

\begin{figure*}[!htbp]
    \subfloat[][Integrable case \label{fig:OTOC_comparison_int}]{
    \includegraphics[width=0.775\columnwidth]{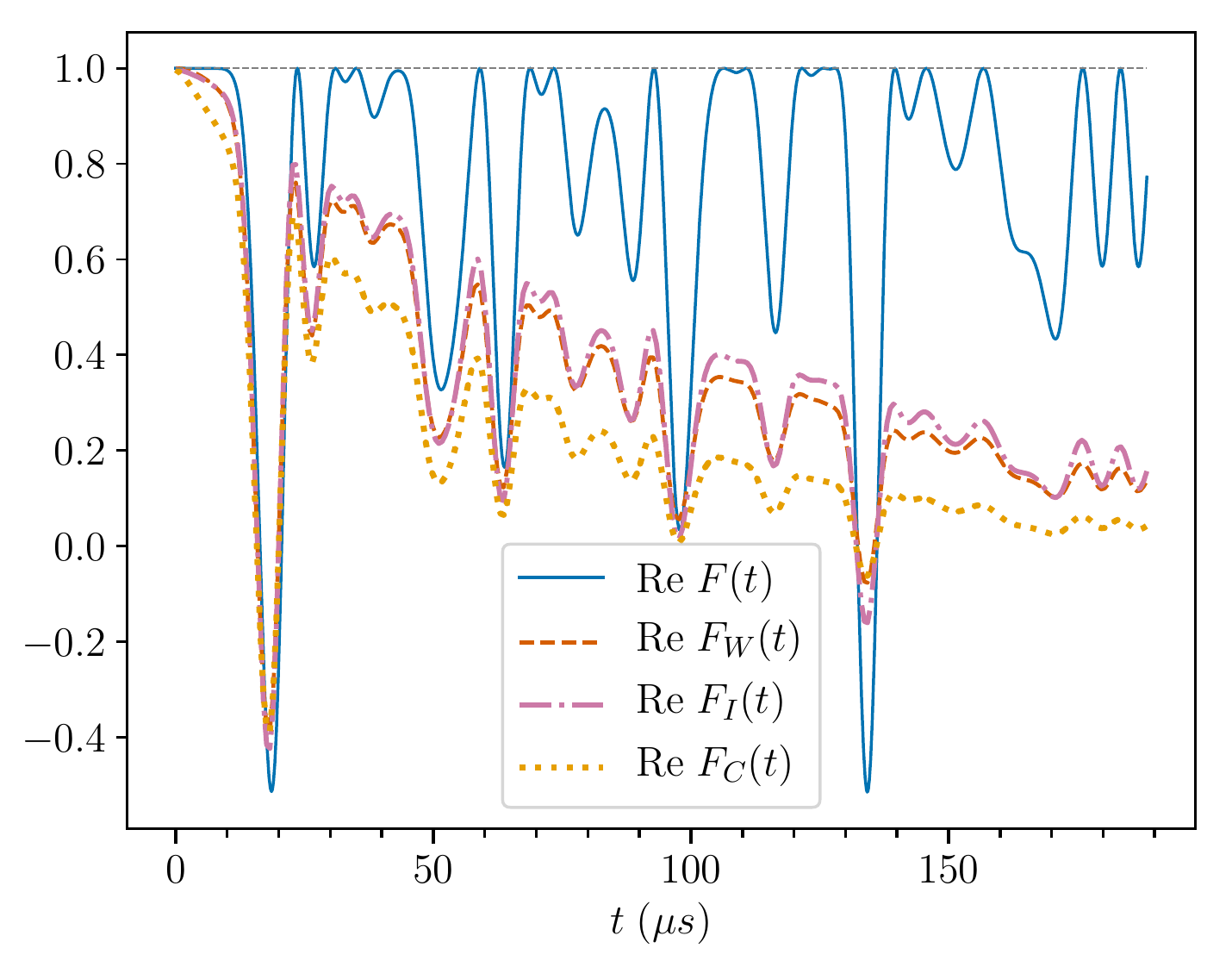}
    }
    \subfloat[Nonintegrable case \label{fig:OTOC_comparison_non-int}]{
    \includegraphics[width=0.775\columnwidth]{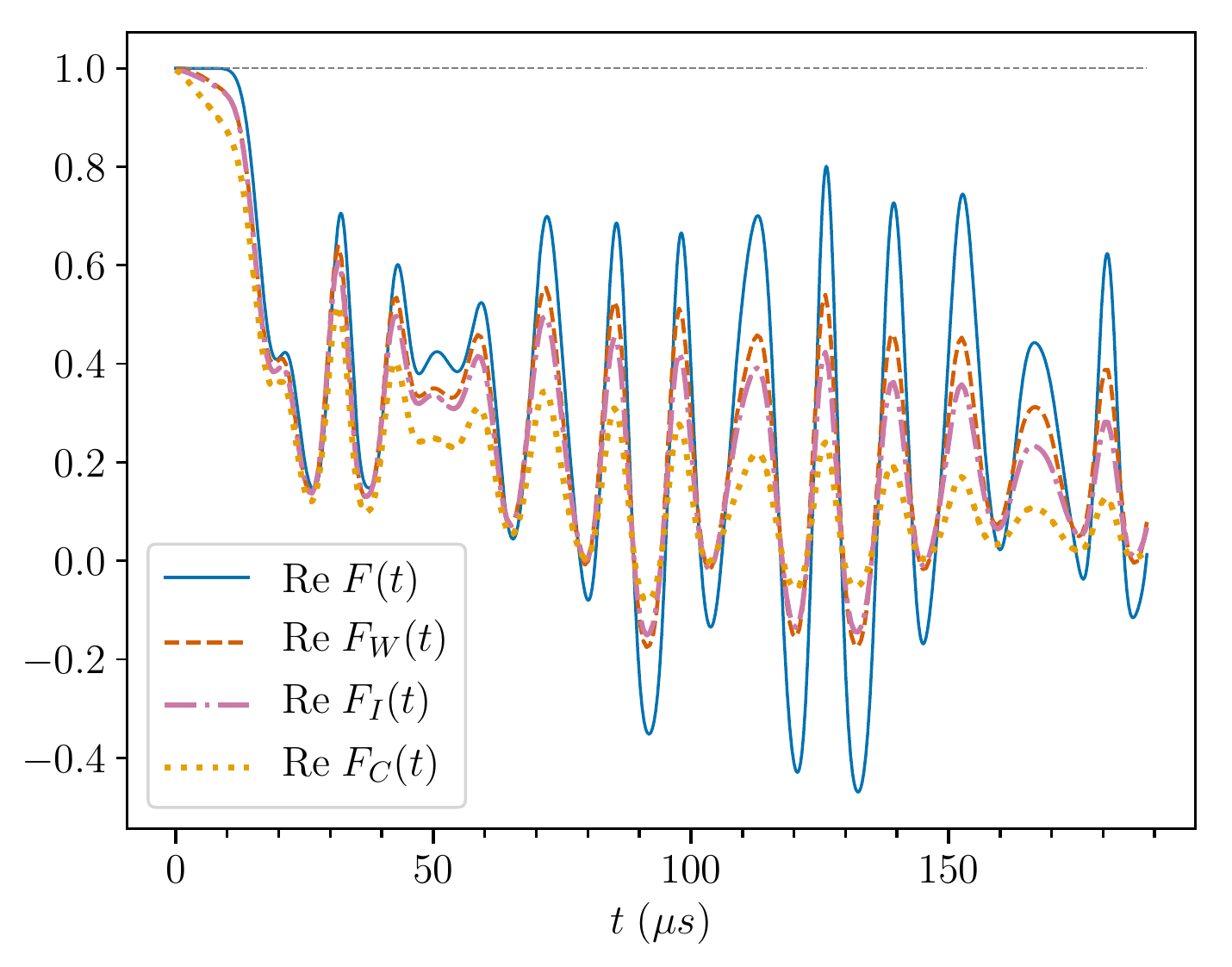}
    }
    \caption{Evolution of measured OTOC,
    $F(t) = \left\langle W^\dagger(t) V^\dagger W(t) V\right\rangle$, with and 
    without decoherence. Values measured with three different protocols are compared against the ideal value: interferometric $F_I (t)$,
    weak $F_W (t)$, and quantum clock $F_C (t)$. To simulate near-term experiments, the system consists of $N=5$ spins in an Ising chain with (a) a transverse field and
    (b) a transverse and a longitudinal field, with parameters detailed in the text. 
    The system starts in a Gibbs state $\rho_T = \mathcal{Z}^{-1}\exp(-H/T)$ with $T/J=1$ and $\mathcal{Z} = \mathrm{Tr}(\exp(-H/T))$.
    The system undergoes environmental dephasing of each qubit with a decay constant of $T_2^*=130\,\mu s$. The local operators
    $W=\sigma_1^z$ and $V=\sigma_N^z$.
    These plots highlight the difficulties in unambiguously distinguishing between (a) nonscrambling and (b) scrambling Hamiltonians in an experimental setting with decoherence.}
    \label{fig:OTOC_comparison}
\end{figure*}
\emph{Decoherence}---We model decoherence with a Lindblad master equation
$d\rho / dt =  -\mathrm{i}[H,\rho] + \sum _{i=1}^{N+n_a}\gamma _{i}\left(L_{i}\rho L_{i}^{\dagger}- 1/2\left\{L_{i}^{\dagger }L_{i},\rho \right\}\right)$.
Here, $N$ denotes the number of spins and $n_a$, the number of ancillas required for a given protocol.
We choose $L_i=\sigma_{i}^z$ and $\gamma_i=\gamma = 1 / ( 2 T_2^*)$. 
The operators $L_i$ implement single-qubit dephasing at rates $\gamma_i$ (dephasing dominates the decoherence).
However, this dephasing also indirectly causes amplitude mixing due to the nondiagonal terms in the Hamiltonian.
The parameter $T_2^*$ denotes the observed exponential decay constant for the qubit coherence from chip-dependent
environmental fluctuations.
We have chosen an optimistic $T_2^* = 130\,\mu s$, plausible for upcoming transmon hardware
\footnote{Private communication with Irfan Siddiqi}.
We interpret the Lindblad equation as an average over the stochastic phase jumps that could occur during each length-$dt$
time step. At each time step, a density matrix $\rho$ updates according to 
\begin{equation}\label{eq:jump_no_jump}
\rho \mapsto dt\sum_i \gamma_i L_{i} U(t)\rho U(t)^\dagger L_{i}^{\dagger} + L_{0} U(t)\rho U(t)^\dagger L_{0}^\dagger.
\end{equation}
The no-phase-jump operator is $L_{0} = \sqrt{\id - dt \sum_i \gamma_i L_{i}^{\dagger} L_{i}}$.
This model offers simplicity and numerical stability \cite{Khezri_Qubitmeasurementerror_2015}.

For each OTOC-measurement protocol, we replace the ideal time evolution with Eq.~\eqref{eq:jump_no_jump} and assume 
that time reversal implements only $U(t)\leftrightarrow U^\dagger(t)$.
We distinguish between the total time elapsed in the laboratory, $t_L$, and the time $t$ at which the OTOC 
is evaluated. Each simulated reversal of $t$ accumulates positive lab time $t_L$; thus, every protocol lasts for a unique $t_L$. To simulate decoherence's effects on the QPD, we 
use the weak-measurement protocol \cite{YungerHalpern_Jarzynskilikeequalityoutoftimeordered_2017,YungerHalpern_QuasiprobabilityOutofTimeOrderedCorrelator_2018}. The other protocols can be adapted for QPD measurements \cite{YungerHalpern_QuasiprobabilityOutofTimeOrderedCorrelator_2018}.

\emph{Simulation results and discussion}---Figure \ref{fig:OTOC_comparison} shows the real part of the OTOC, measured in the presence of decoherence: 
$F_{I}(t)$, $F_{W}(t)$, and $F_{C}(t)$ denote the OTOC measured according to the interferometric \cite{Swingle_Measuringscramblingquantum_2016},
weak-measurement \cite{YungerHalpern_Jarzynskilikeequalityoutoftimeordered_2017,YungerHalpern_QuasiprobabilityOutofTimeOrderedCorrelator_2018}, and quantum-clock \cite{Zhu_Measurementmanybodychaos_2016} 
protocols \footnote{See Supplemental Material for details on how to numerically simulate each OTOC-measurement protocol.}. These curves are compared to the ideal OTOC $F(t)$ measured in the absence 
of noise.
These protocols differ in the amounts of lab time required to measure $F(t)$: the protocols need $t_L$'s that are at least  $2t$, $3t$, and $4t$, respectively. As expected, OTOCs measured with long-$t_L$ protocols decay the most, since they suffer from decoherence the longest. The quantum-clock protocol's
$F_C(t)$ is affected the most. Nonetheless, this protocol's essence---the implementation of time reversals via an ancilla qubit---could
be combined with a shorter-$t_L$ protocol (e.g., the interferometric protocol), to mitigate decoherence \cite{Dressel_StrengtheningWeakMeasurements_2018}.

Figures \ref{fig:OTOC_comparison_int} and \ref{fig:OTOC_comparison_non-int} show that decoherence hinders us from easily distinguishing 
between integrable and nonintegrable Hamiltonians. The integrable-Hamiltonian OTOC with decoherence decays due to information leaking, 
and the nonintegrable-Hamiltonian OTOC revives.
If we used these two OTOCs' qualitative behaviors, we would misclassify the Hamiltonians and incur a false positive, inferring scrambling 
where there is none.

\begin{figure*}[!htbp]
	\subfloat[Integrable ideal case \label{fig:quasi_comparison_int_ideal}]{
        \includegraphics[width=0.775\columnwidth]{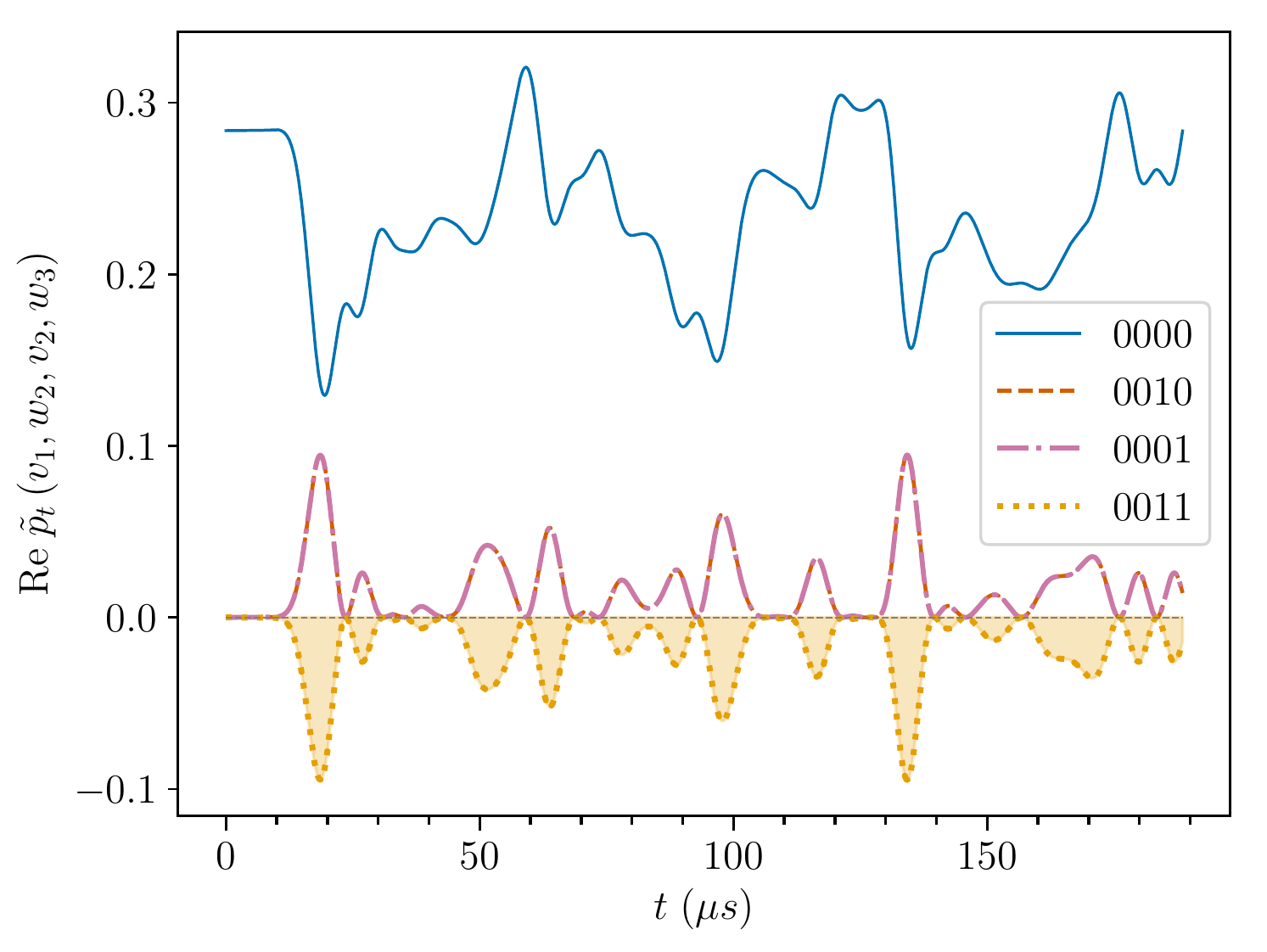}
    }
    \subfloat[Integrable case with decoherence \label{fig:quasi_comparison_int_decoh}]{
        \includegraphics[width=0.775\columnwidth]{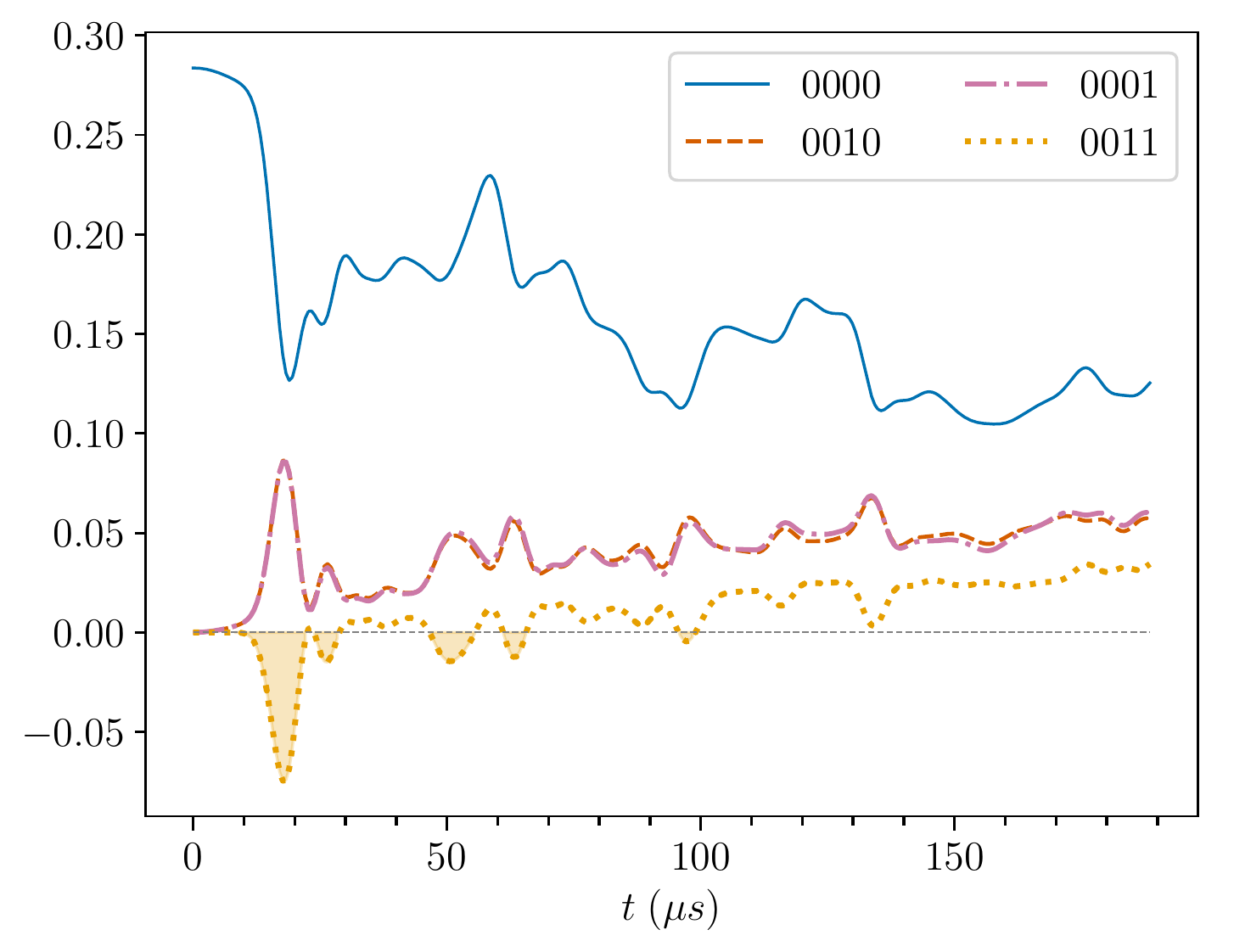}
    }\\
    \subfloat[Nonintegrable ideal case \label{fig:quasi_comparison_non-int_ideal}]{
        \includegraphics[width=0.775\columnwidth]{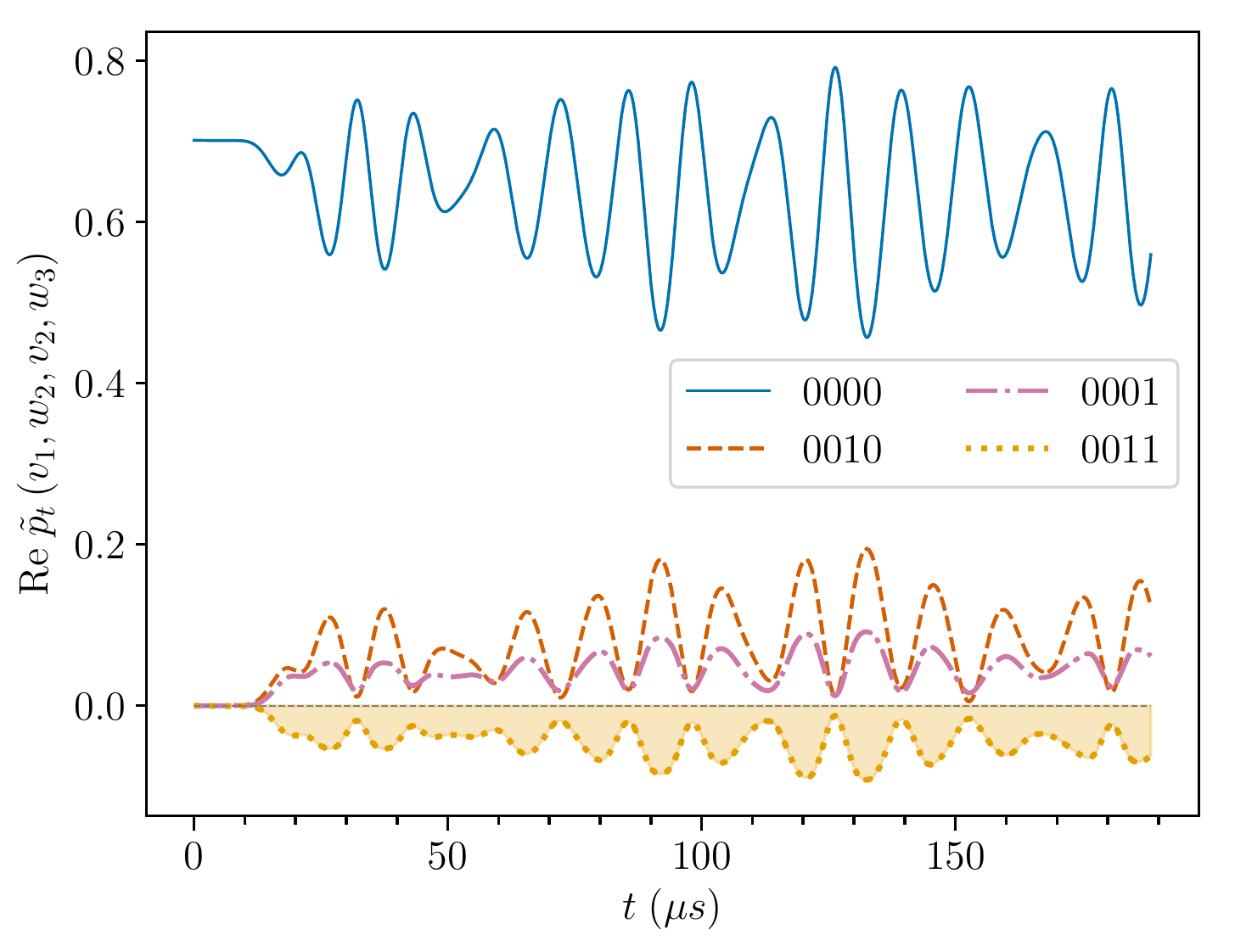}
    }
    \subfloat[Nonintegrable case with decoherence \label{fig:quasi_comparison_non-int_decoh}]{
        \includegraphics[width=0.775\columnwidth]{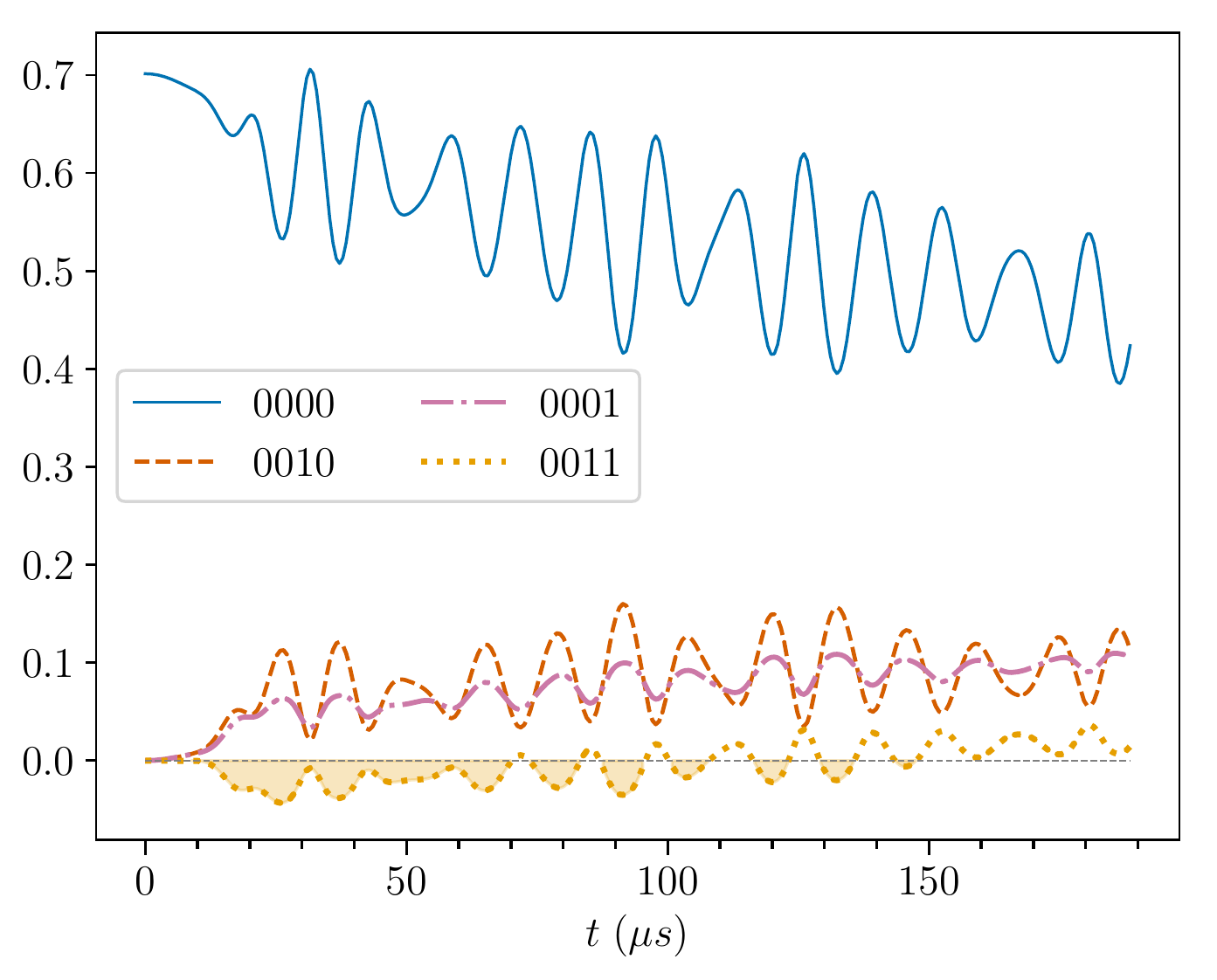}
    }
    \caption{Evolution of measured $\mathrm{Re}\;\tilde{p}_t$ with and without decoherence, using the sequential-weak-measurement protocol.
    The QPD, $\tilde{p}_t(v_1, w_2, v_2, w_3) = \mathrm{Tr}(\Pi_{w_3}^{W(t)} \Pi_{v_2}^{V} \Pi_{w_2}^{W(t)} \Pi_{v_1}^{V} \rho)$, underlies the OTOC, $F(t) = \sum v_1 w_2 v_2^* w_3^*\,\tilde{p}_t(v_1, w_2, v_2, w_3)$, where $V = \sum v \Pi_v$ and $W = \sum w \Pi_w$.
    Of the sixteen QPD values, four examples are shown. The numeric labels in the legend have the form $abcd$,
    where $v_1=(-1)^a$, $w_2=(-1)^b$, $v_2=(-1)^c$, and $w_3=(-1)^d$. The shaded regions show nonclassical behavior of the QPD.}
    \label{fig:quasi_comparison}
\end{figure*}

Distinguishing scrambling from integrable Hamiltonians via the QPD is straightforward, despite decoherence (Fig.~\ref{fig:quasi_comparison}). Decoherence damps the distribution's oscillations, and the different curves drift 
towards a common value (in our example, between 0 and 0.1). Unlike in the integrable case, the nonintegrable case's 
quasiprobability shows a persistent bifurcation that we call a pitchfork: around $t\approx 15\mu s$ quasiprobabilities that used to lie atop each at $y=0$ split. This pitchfork arises because scrambling breaks a symmetry as it eliminates the QPD’s invariance under certain permutations and negations of the QPD arguments in Eq.~\eqref{eq:coarse_quasi_prob}
\cite{YungerHalpern_QuasiprobabilityOutofTimeOrderedCorrelator_2018}.
The symmetry breaking eliminates the QPD’s constancy under certain interchanges, and certain negations,
of measurement outcomes in a weak-measurement trial. We should expect this asymmetry to surface in the total nonclassicality $\tilde{N}_t$ 
of Eq.~\eqref{eq:total_non-class}. Since information scrambling is related to many-body entanglement, which is nonclassical,
we expect the QPD's nonclassicality to be a robust indicator of scrambling. Indeed, damping shrinks
the negative regions in Fig.~\ref{fig:quasi_comparison}. The negative regions also show structure that mirrors qualitative behavior of the OTOC:
the decay of $\mathrm{Re}\; F(t)$ matches the flourishing of the negativity; the revivals of $\mathrm{Re}\; F(t)$ mirror the negativity's disappearance. Yet the QPD provides information absent from $F(t)$.

\begin{figure*}[!htbp]
        \subfloat[Integrable case \label{fig:abs_comparison_int}]{
           \includegraphics[width=0.78\columnwidth]{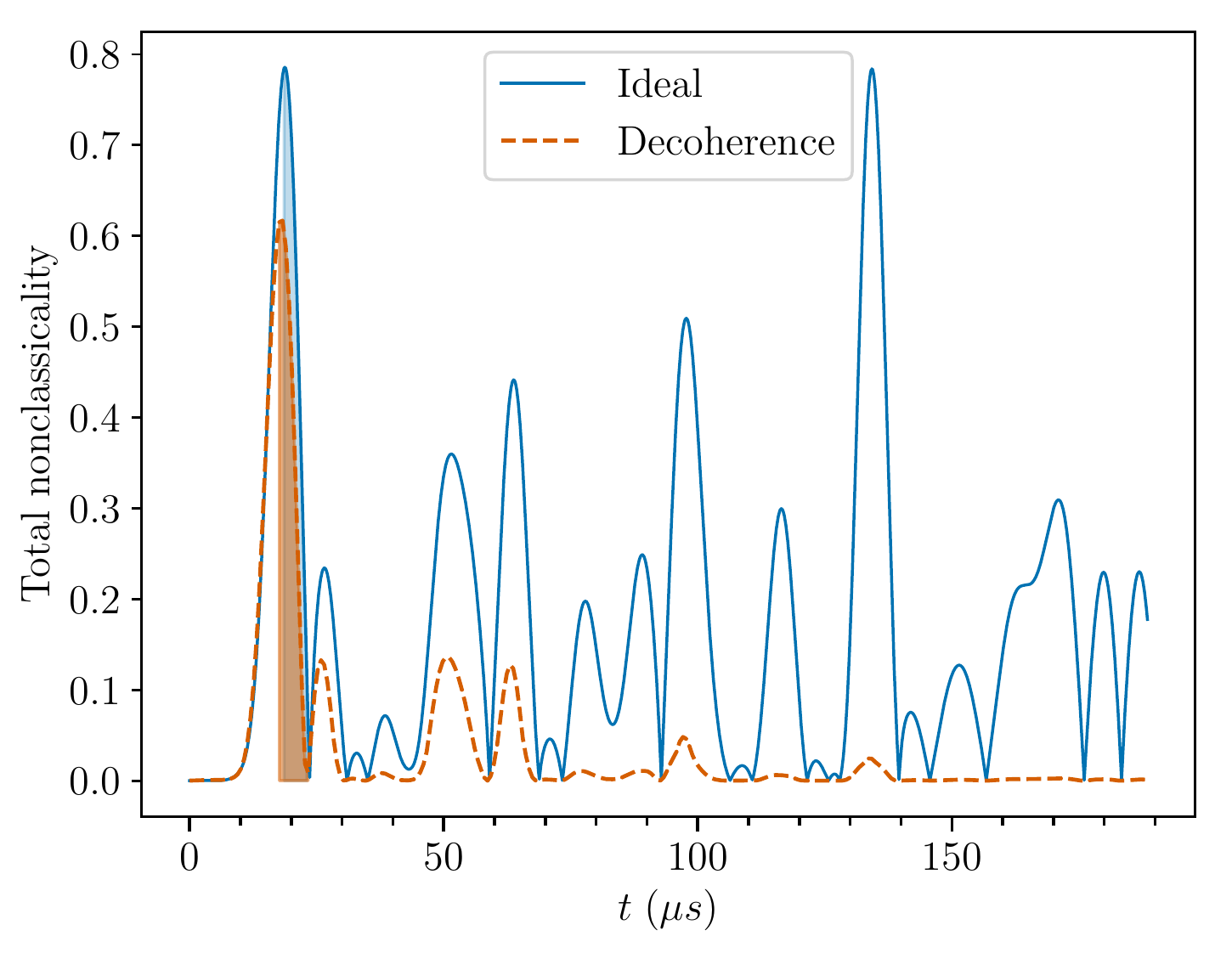} 
        }
        \subfloat[Nonintegrable case \label{fig:abs_comparison_non-int}]{
            \includegraphics[width=0.78\columnwidth]{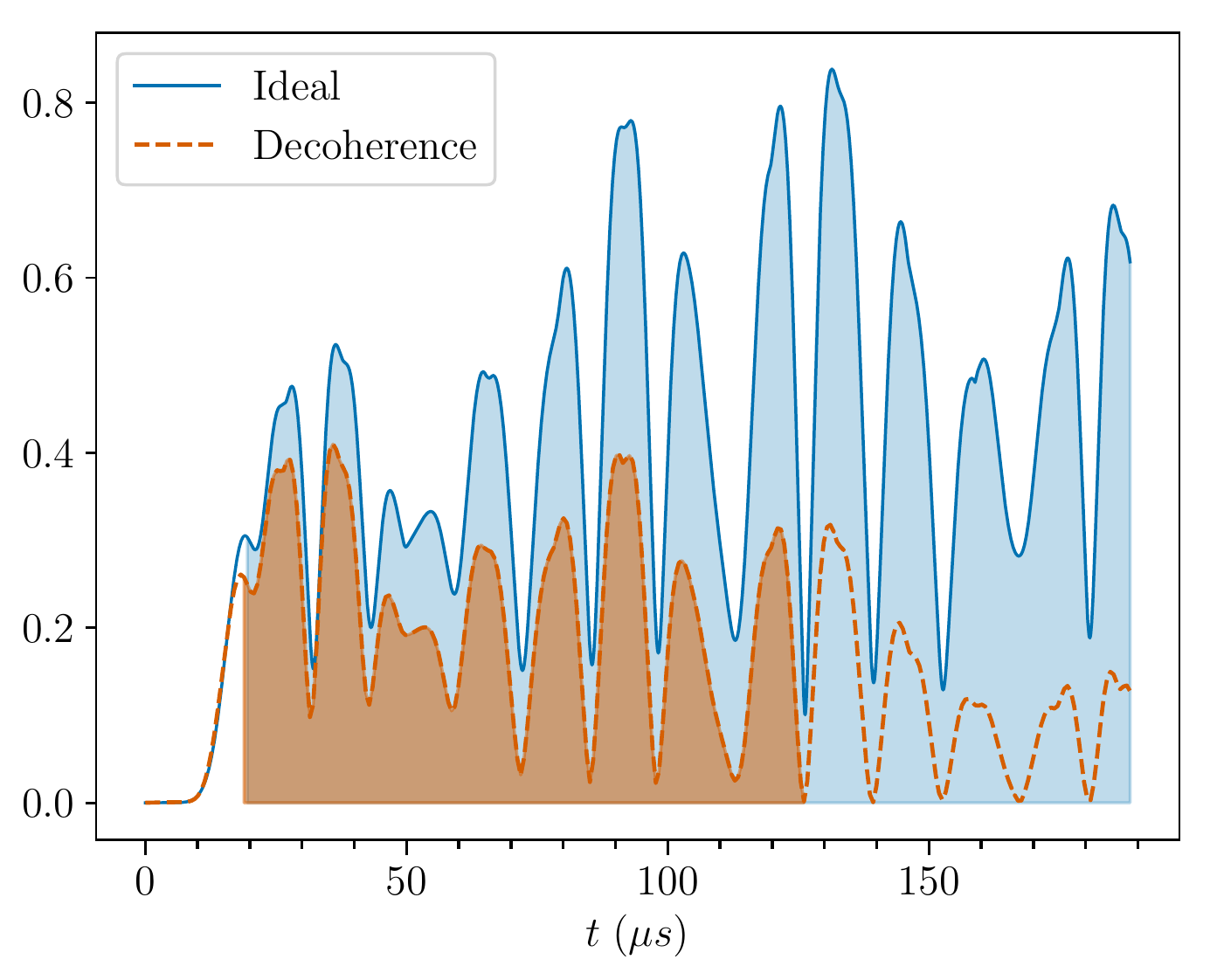}
        }
    \caption{Total nonclassicality, $\tilde{N}(t) = \sum |\tilde{p}_t(v_1, w_2, v_2, w_3)| - 1$, of the QPD, $\tilde{p}_t$, showing sensitivity to 
    decoherence for (a) integrable and (b) scrambling systems. Comparing two timescales can reveal scrambling. The duration between the onset of 
    nonclassicality ($\tilde{t}_* \sim 10\,\mu s$) and the first maximum ($t_m \sim 20\,\mu s$) is roughly constant across both plots. The
    area between $t_m$ and the next zero ($t_z$) is shaded. For the integrable Hamiltonian, 
    $t_z-t_m \sim t_m-\tilde{t}_* \sim 10\,\mu s$. For the nonintegrable Hamiltonian, $t_z-t_m$ remains an order of magnitude larger ($t_z-t_m \sim 100\,\mu s$), even with decoherence. In the decoherence-free scrambling case, $\tilde{N}(t)$ remains nonzero for at least four orders of magnitude of time longer than in the nonscrambling case.} \label{fig:abs_comparison}
\end{figure*}

We plot $\tilde{N}(t)$ in Fig.~\ref{fig:abs_comparison}. The nonclassicality's persistence reflects sustained noncommutativity of $W(t)$ and $V$.
Denote by $\tilde{t}_*$ the point at which $\tilde{N}(t)$ first deviates
from zero \footnote{With this definition, $\tilde{t}_*$ is close to the scrambling time $t_*$ at which
$\mathrm{Re}\;F(t)$ first deviates significantly from one.}; by $t_m$, the point at which the first maximum occurs; and by $t_z$, the 
time at which the first subsequent zero happens.
For the scrambling dynamics with decoherence in Fig.~\ref{fig:abs_comparison}, $t_z-t_m$ is more than an order
of magnitude longer than $t_m-\tilde{t}_*$. For the nonscrambling dynamics, the two timescales are comparable. In this case, and without 
dissipation, $t_z-t_m$ is longer than the simulation time. We thus conjecture that, if $t_m - \tilde{t}_* \ll t_z - t_m$,
the dynamics is scrambling \footnote{See Supplemental Material for details on how the total non classicality and its time scales vary with $h=J$.}. As quantum information spreads throughout the system in a time $t_m-\tilde{t}_*\propto N$,
if $H$ is integrable, some information recollects in a time $t_z-t_m\propto N$. Hence, the total nonclassicality's
first peak should be approximately symmetrical. If the system dynamics is scrambling, such a recollection would occur after a much longer time \cite{Bocchieri_QuantumRecurrenceTheorem_1957,CamposVenuti_recurrencetimequantum_2015,Hosur_Chaosquantumchannels_2016}.
$\tilde{N}(t)$ should display strong temporal asymmetry about its first maximum. We see this lack of symmetry in the scrambling case's 
$\tilde{N}(t)$ in Fig.~\ref{fig:abs_comparison_non-int}.

We see also our conjecture's role in the presence of decoherence: because of
the significant differences in the scrambling-case timescales, the asymmetry persists despite the dissipation's suppression of $\tilde{N}(t)$.
$F(t)$ offers no such quantitative insight: $\tilde{N}(t)$ is useful because it precisely identifies when nonclassical behavior arises and disappears.

\emph{Conclusions and outlook}---We propose that a more robust witness can be found in the nonclassical part of the QPD $\tilde{p}_t$
behind the OTOC. The total nonclassicality $\tilde{N}$ of $\tilde{p}_t$ helps distinguish 
integrable from scrambling Hamiltonians in the presence of decoherence. One can distinguish clearly
between scrambling and nonscrambling systems by comparing two timescales of $\tilde{N}$.
The duration between the birth of nonclassicality, at the time $\tilde{t}_*$, and the nonclassicality's
first local maximum, at $t_m$, is related to the time needed by quantum information to spread throughout
the system. The spreading's persistence governs the duration between $t_m$ and the
death of nonclassicality, at $t_z$. Nonscrambling dynamics exhibit revivals of
classicality on timescales $t_m - \tilde{t}_* \approx t_z - t_m$, while scrambling dynamics take much longer.
This distinction is seen clearly in the total nonclassicality $\tilde{N}(t)$. Unlike the OTOC,
$\tilde{N}(t)$ is robust with respect to experimental imperfections like decoherence.
Characterizing this time’s scaling with system size, and checking whether the scaling can be consistent with doubly exponential
expectations inspired by the Poincaré recurrence time \cite{Bocchieri_QuantumRecurrenceTheorem_1957,CamposVenuti_recurrencetimequantum_2015,Hosur_Chaosquantumchannels_2016}, is a subject for future research.

This study of decoherence highlights two opportunities for improving the robustness and convenience of the QPD-measurement scheme in \cite{YungerHalpern_QuasiprobabilityOutofTimeOrderedCorrelator_2018}.
First, the weak measurements' coupling might be strengthened, along the lines in \cite{Dressel_StrengtheningWeakMeasurements_2018}.
Second, the scheme in \cite{Swingle_Resiliencescramblingmeasurements_2018} might be applied to renormalize away experimental errors.

Another opportunity for future study is whether scrambling breaks symmetries in OTOC QPDs defined in terms of $W$ and $V$ operators
other than qubit Pauli operators. An interesting choice to study next would be the Sachdev-Ye-Kitaev (SYK) model \cite{Sachdev_Gaplessspinfluidground_1993,Kitaev_SimpleModelQuantum_2015}. The SYK model consists of Majorana fermions, whose experimental
realizations are being pursued assiduously 
\cite{Hassler_MajoranaQubits_2014,Aasen_MilestonesMajoranabasedquantum_2016,Deng_MajoranaNonlocalityHybrid_2018,Vaitiekenas_EffectiveGfactorMajorana_2018,Lutchyn_RealizingMajoranaZero_2018,Gomez-Ruiz_Universaltwotimecorrelations_2018,OBrien_MajoranaBasedFermionicQuantum_2018}. As the SYK model scrambles maximally quickly, like black holes, it has been hoped to shed light on 
quantum gravity. The calculational tools available for SYK merit application to the OTOC QPD, which may shed new light on scrambling at the 
intersection of condensed matter and high-energy physics.

\begin{acknowledgments}
\emph{Acknowledgements}---JRGA was supported by a fellowship from the Grand Challenges Initiative at Chapman University.
NYH is grateful for funding from the Institute for Quantum Information and Matter, an NSF Physics Frontiers Center (NSF 
Grant PHY-1125565) with support from the Gordon and Betty Moore Foundation (GBMF-2644); a Graduate Fellowship from the 
Kavli Institute for Theoretical Physics, supported by the NSF under Grant No. NSF PHY-1125915; the Walter Burke Institute 
for Theoretical Physics at Caltech; and a Barbara Groce Graduate Fellowship.
JD was partially supported by the Army Research Office (ARO) Grants No. W911NF- 300 15-1-0496 and No. W911NF-1-81-0178.
The authors wish to thank Paul Dieterle, Poul Jessen, Andrew Keller, Oskar Painter, and Mordecai Waegell for helpful discussions.
\end{acknowledgments}

\bibliography{../references/Reference_Database}
\end{document}


\title{Supplemental material: Out-of-time-ordered-correlator quasiprobabilities robustly witness scrambling}
\author{José Raúl González Alonso}
\email[Corresponding author: ]{gonzalezalonso@chapman.edu}
\affiliation{Schmid College of Science and Technology, Chapman University, Orange, CA 92866, USA}
\author{Nicole Yunger Halpern}
\affiliation{Institute for Quantum Information and Matter, Caltech, Pasadena, CA 91125, USA}
\author{Justin Dressel}
\affiliation{Schmid College of Science and Technology, Chapman University, Orange, CA 92866, USA}
\affiliation{Institute for Quantum Studies, Chapman University, Orange, CA 92866, USA}

\date{\today}
\maketitle

\setcounter{equation}{0}
\setcounter{figure}{0}
\setcounter{table}{0}
\setcounter{page}{1}
\renewcommand{\theequation}{S\arabic{equation}}
\renewcommand{\thefigure}{S\arabic{figure}}
\renewcommand{\bibnumfmt}[1]{[S#1]}
\renewcommand{\citenumfont}[1]{S#1}

\section{OTOC Measurement Protocols and Numerical Simulations}

Here, we provide more details about the numerical procedure used to simulate each OTOC-measurement protocol. 
For each protocol, the time-evolution implementation changes, depending on whether we are simulating a closed or an open system.
In the case of a closed system, the forward time evolution amounts to applying the unitary $U(t)$ generated by the appropriate Hamiltonian.
For the open system, we evolve a given operator $O$ by
\begin{equation}
O \mapsto dt\sum_i \gamma_i L_{i} U(dt)O U(dt)^\dagger L_{i}^{\dagger} + L_{0} U(dt) O U(dt)^\dagger L_{0}^\dagger.
\end{equation}
where $L_{0} = \sqrt{\id - dt \sum_i \gamma_i L_{i}^{\dagger} L_{i}}$ and the $L_i$'s are the Lindblad operators associated with the decoherence.
In our case, the Lindblad operators represent individual-qubit dephasing. The backward time evolution flips the sign of $t$ only for the unitary
evolution, i.e., $U(t)\leftrightarrow U^\dagger(t)$. This method ensures state positivity, unlike more direct methods for integrating the master
equation. It is also possible to construct a superoperator matrix to simulate 
the dynamics of the system, even if decoherence is present. However, because of memory constraints, and the lack of symmetries with which to simplify
the problem, this method becomes impractical even for a modest number of qubits.

In what follows, we detail the numerical procedures used to simulate the measurement protocols outlined in Refs. \cite{YungerHalpern_Jarzynskilikeequalityoutoftimeordered_2017,YungerHalpern_QuasiprobabilityOutofTimeOrderedCorrelator_2018,Swingle_MeasuringScramblingQuantum_2016,Zhu_MeasurementManybodyChaos_2016}. Note that in our numerical procedures we only simulate the decoherence accumulated in each protocol but, for simplicity, ignore other experimental imperfections.
Obtaining a graph over a range of $t$ values for an OTOC simulation is not a simple Lindblad integration over a set of stored time points, since computing an OTOC for one time point $t$ corresponds to multiple legs of forward and backward evolution, depending on the simulated protocol. This folded evolution structure necessitates a complete integration of duration several $t$ for each plotted $t$ value, and prior integrations may not be used to compute later $t$ values. In light of these challenges, and to target near-term experiments, we have chosen to simulate $N=5$ qubits. This number of qubits is sufficient to observe the onset of many-body scrambling behavior while remaining experimentally feasible.

\subsection{Weak-measurement protocol}
To calculate $\Tr\left( A B C D \rho \right)$, we implement the procedure outlined below, based on \cite{YungerHalpern_Jarzynskilikeequalityoutoftimeordered_2017,YungerHalpern_QuasiprobabilityOutofTimeOrderedCorrelator_2018}. One advantage of the
weak-measurement protocol is that we can use it to calculate either the OTOC quasiprobability 
 $\tilde{p}_t (v_1, w_2, v_2, w_3)$ (by using the projectors onto eigenspaces,
 $A=\Pi_{w_3}^{W}, B=\Pi_{v_2}^{V}, C=\Pi_{w_2}^{W},$ and $D=\Pi_{v_1}^{V}$) or the OTOC $F(t)$ (with $A=W(t)^\dagger, B=V^\dagger, C=W(t),$ and $D=V$). The unitary evolution is generated by the system Hamiltonian tensored with identity operators on the ancillas required by the protocol.
 The steps in the calculation are as follows:
\begin{enumerate}
\item Prepare $\rho$.
\item Left-multiply by $D$.
\item Evolve the result forward in time by $t$ units.
\item Left-multiply the result by $C$.
\item Evolve the result backward in time by $t$ units.
\item Left-multiply the result by $B$.
\item Evolve the result forward in time by $t$ units.
\item Left-multiply the result by $A$, and take the trace to obtain $F(t)$ or $\tilde{p}_t (v_1, w_2, v_2, w_3)$.
\end{enumerate}
While taking the trace is a trivial operation in theory, in an experiment, it is necessary to repeat the procedure outlined above 
multiple times and calculate the average of the outcomes, using their relative frequencies to then obtain the trace. Similarly, a left or right
multiplication involves a weak coupling to an ancilla followed by the measurement of an appropriate observable on the ancilla. For explicit examples on how to do this with qubits see Ref. \cite{Dressel_StrengtheningWeakMeasurements_2018}.

\subsection{Interferometric protocol}
The Swingle \textit{et. al.} interferometric protocol \cite{Swingle_MeasuringScramblingQuantum_2016} uses an ancilla to apply different
operators selectively.
On one branch of an interferometer the product of the operators $V$ and $W(t)$ is applied. Meanwhile, on the other branch, the product in
reversed order is applied. 
The unitary evolution is generated by the system Hamiltonian tensored with an identity operator on the ancilla that creates the interferometric
branching. We have used the following numerical procedure:
\begin{enumerate}
\item Prepare $\rho\otimes\ketbra{+}{+}$.
\item Apply $\id\otimes\ketbra{0}{0} + V \otimes \ketbra{1}{1}$.
\item Evolve the result forward in time by $t$ units.
\item Apply $W \otimes\id$ to the result.
\item Evolve the result backward in time by $t$ units.
\item Apply $V \otimes\ketbra{0}{0} + \id\otimes\ketbra{1}{1}$.
\item Measuring the control qubit in the $\sigma_x$ or the $\sigma_y$ eigenbasis yields $\mathrm{Re}\, F(t)$ or $\mathrm{Im}\,F(t)$ respectively.
\end{enumerate}

\subsection{Quantum-clock protocol}
Finally, the Zhu \textit{et. al.} quantum-clock protocol \cite{Zhu_MeasurementManybodyChaos_2016}, like the Swingle \textit{et. al.}
interferometric protocol, relies on an ancilla to selectively apply the product of $W$ and $V$ in different orders to each branch. However,
the unitary evolution is generated by the system Hamiltonian $H$ tensored with a $\sigma_z$ on said ancilla, i.e., the total Hamiltonian is
$H_{T} = H\otimes\sigma_z$. The advantage of this procedure is that the ancilla also controls the direction of time evolution.
In other words, $H_T$ generates a unitary of the form
\begin{equation}
U_T(t) = U(t) \otimes \ketbra{0}{0}  + U(-t) \otimes \ketbra{1}{1}.
\end{equation}
The numerical procedure we used in this case is outlined below:
\begin{enumerate}
\item Prepare $\rho\otimes\ketbra{+}{+}$
\item Apply $\id\otimes\ketbra{0}{0} + V \otimes\ketbra{1}{1}$.
\item Evolve in time by $t$ units.
\item Apply $\id\otimes\ketbra{0}{0} + W \otimes\ketbra{1}{1}$.
\item Apply $\id\otimes \sigma_x$.
\item Evolve in time by $2t$ units.
\item Apply $\id\otimes \sigma_x$.
\item Apply $W \otimes\ketbra{0}{0} + \id \otimes\ketbra{1}{1}$.
\item Evolve in time by $t$ units.
\item Apply $V\otimes\ketbra{0}{0} + \id \otimes\ketbra{1}{1}$.
\item Measuring the clock qubit in the $\sigma_x$ or the $\sigma_y$ eigenbasis yields $\mathrm{Re}\, F(t)$ or $\mathrm{Im}\,F(t)$, respectively. 
\end{enumerate}

\section{Changes in the behaviors of \texorpdfstring{$\tilde{t}_*$}{t*}, \texorpdfstring{$t_m$}{tm}, and \texorpdfstring{$t_z$}{tz} as \texorpdfstring{$h/J$}{h over J} varies}

In the main text, we analyzed the behaviors of three different time scales in the total nonclassicality $\tilde{N}(t)$ of the
quasiprobability $\tilde{p}_t$, for two values of $h/J$. All these time scales are analyzed up to numerical imprecisions
given by the square of the time step $\Delta t$ used in the simulations. The definitions of the time scales are given below:
\begin{enumerate}
    \item $\tilde{t}_*$, the time at which $\tilde{N}(t)$ first deviates from zero.
    \item $t_m$, the time at which $\tilde{N}(t)$ attains its first local maximum.
    \item $t_z$, the time at which $\tilde{N}(t)$ first returns to zero after the first maximum.
\end{enumerate}
We saw that, even in the presence of decoherence, the asymmetry between $t_m-\tilde{t}_*$ and $t_z-t_m$ distinguished between the integrable
$(h/J =0)$ and nonintegrable ($h/J\ne 0$) cases.
In the plots below, we analyze the behaviors of the time scales for 15 equally spaced values of $h/J$ between $0.0$ and $0.5$.

\begin{figure*}[htbp]
        \subfloat[Ideal case \label{fig:tot_non_class_inf_noiseless_comp}]{
           \includegraphics[width=0.45\linewidth]{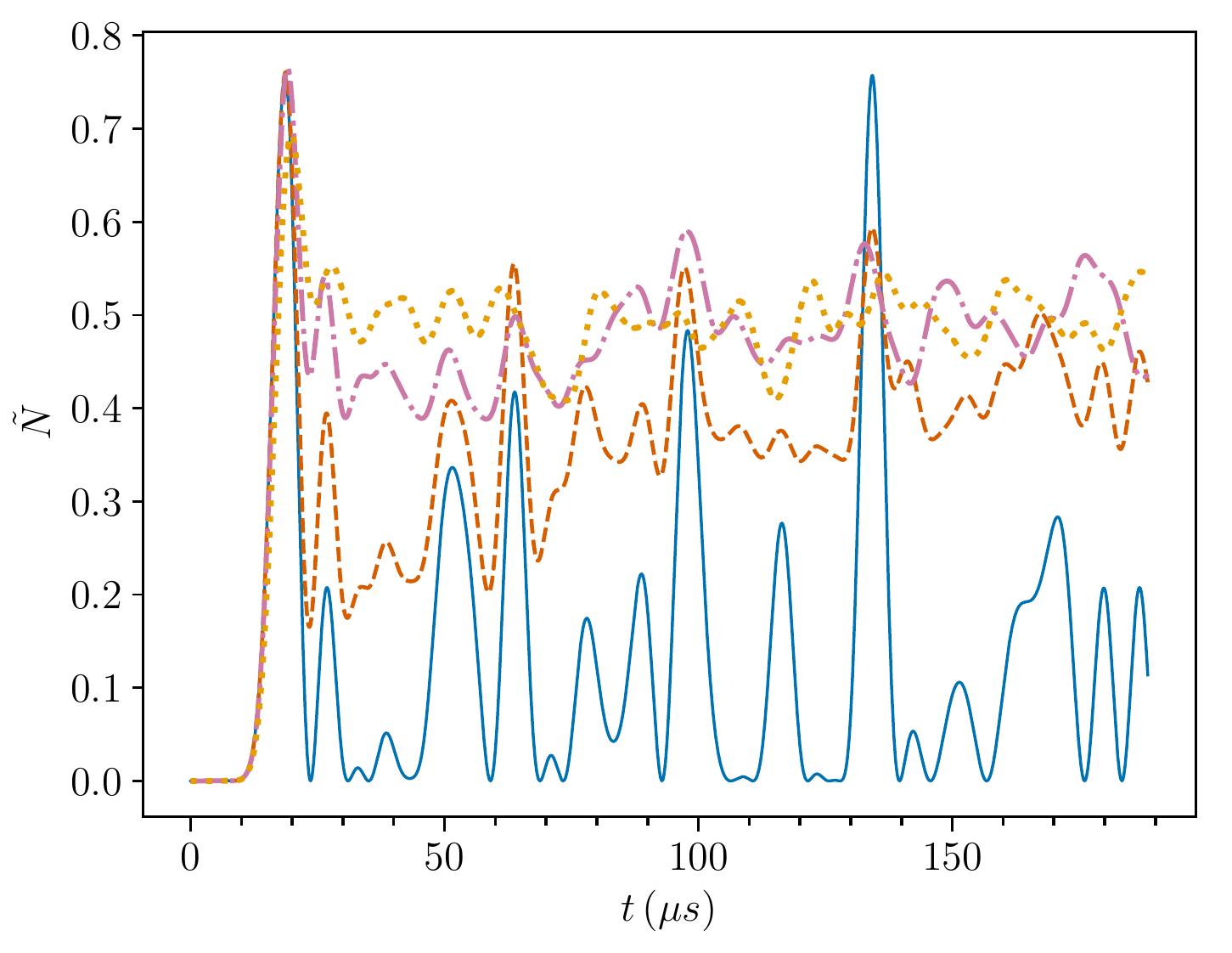} 
        }
        \subfloat[Decoherent case \label{fig:tot_non_class_inf_noisy_comp}]{
            \includegraphics[width=0.45\linewidth]{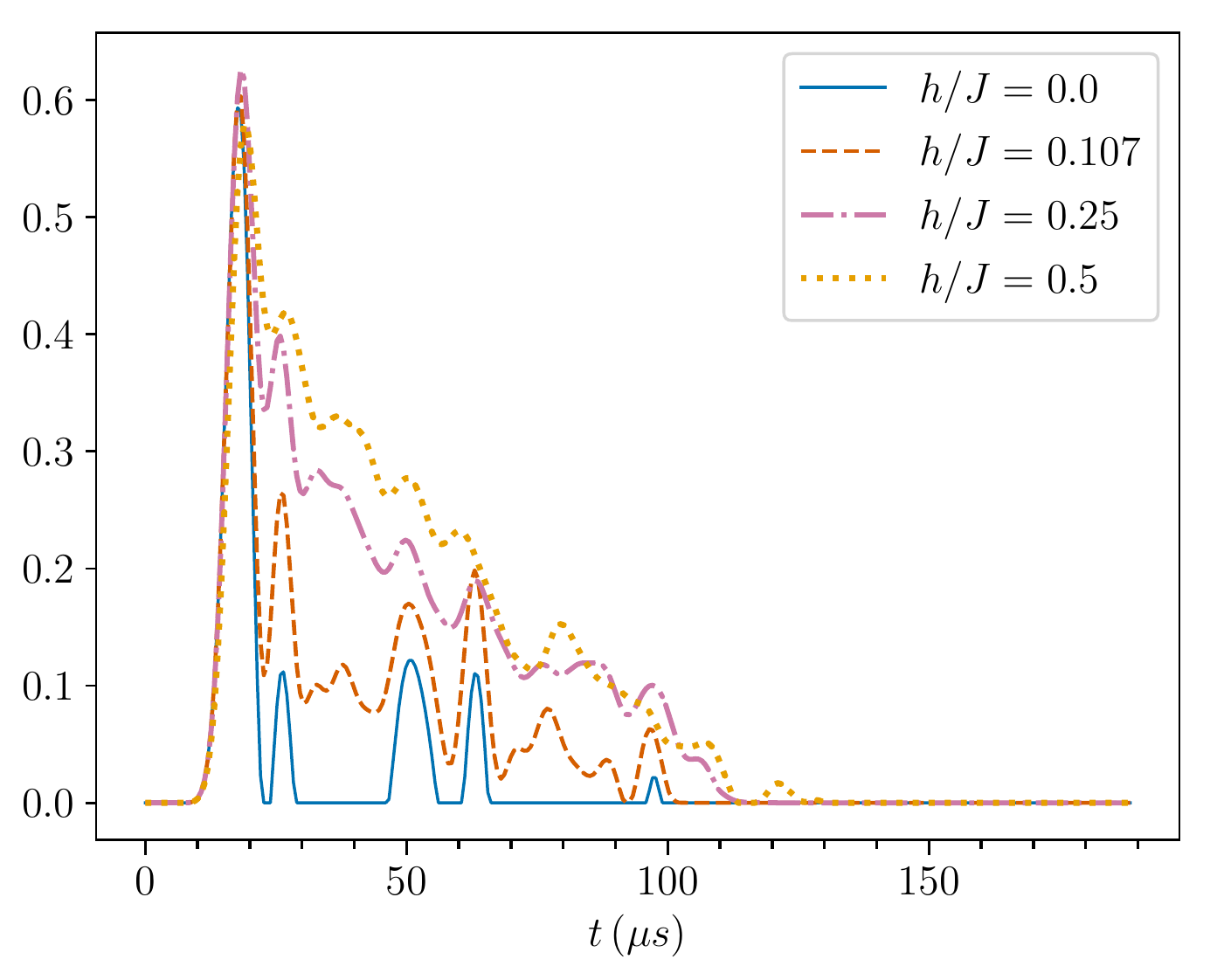}
        }
    \caption{Behavior of $\tilde{N}(t)$ for different values of $h/J$ for the (a) ideal and (b) decoherent cases with an initial
    infinite-temperature Gibbs state, $\id/2^N$. In the decoherent case, the system undergoes environmental dephasing of each qubit with a decay constant
    of $T^*_2 = 130\;\mu s$.
    The local operators are $W = \sigma_1^z$ and $V = \sigma_N^z$. These plots highlight how $h/J$ controls not only integrability and 
    scrambling, but also cumulative nonclassicality.}\label{fig:tot_non_class_inf_comp}
\end{figure*}

First, we present an example of the effects that changing $h/J$ has on the total nonclassicality $\tilde{N}(t)$. As we can see in 
Fig.~\ref{fig:tot_non_class_inf_comp}, as the value of $h/J$ increases, so does the cumulative total nonclassicality. This feature
is independent of decoherence. Hence, we can think of $h/J$ as a parameter that controls not only the scrambling nature of the 
Hamiltonian but also the cumulative behavior of the total nonclassicality $\tilde{N}(t)$. For instance, we see a sharp transition
in $t_z$ at $h/J=0$: $\tilde{N}(t)$ for the integrable case promptly returns to zero after its first maximum but takes longer for all
the nonintegrable cases. This behavior is expected since $h/J=0$ indicates when the system is integrable.

\begin{figure*}[htbp]
        \subfloat[Infinite temperature \label{fig:ratio_inds_inf_comp}]{
           \includegraphics[width=0.45\linewidth]{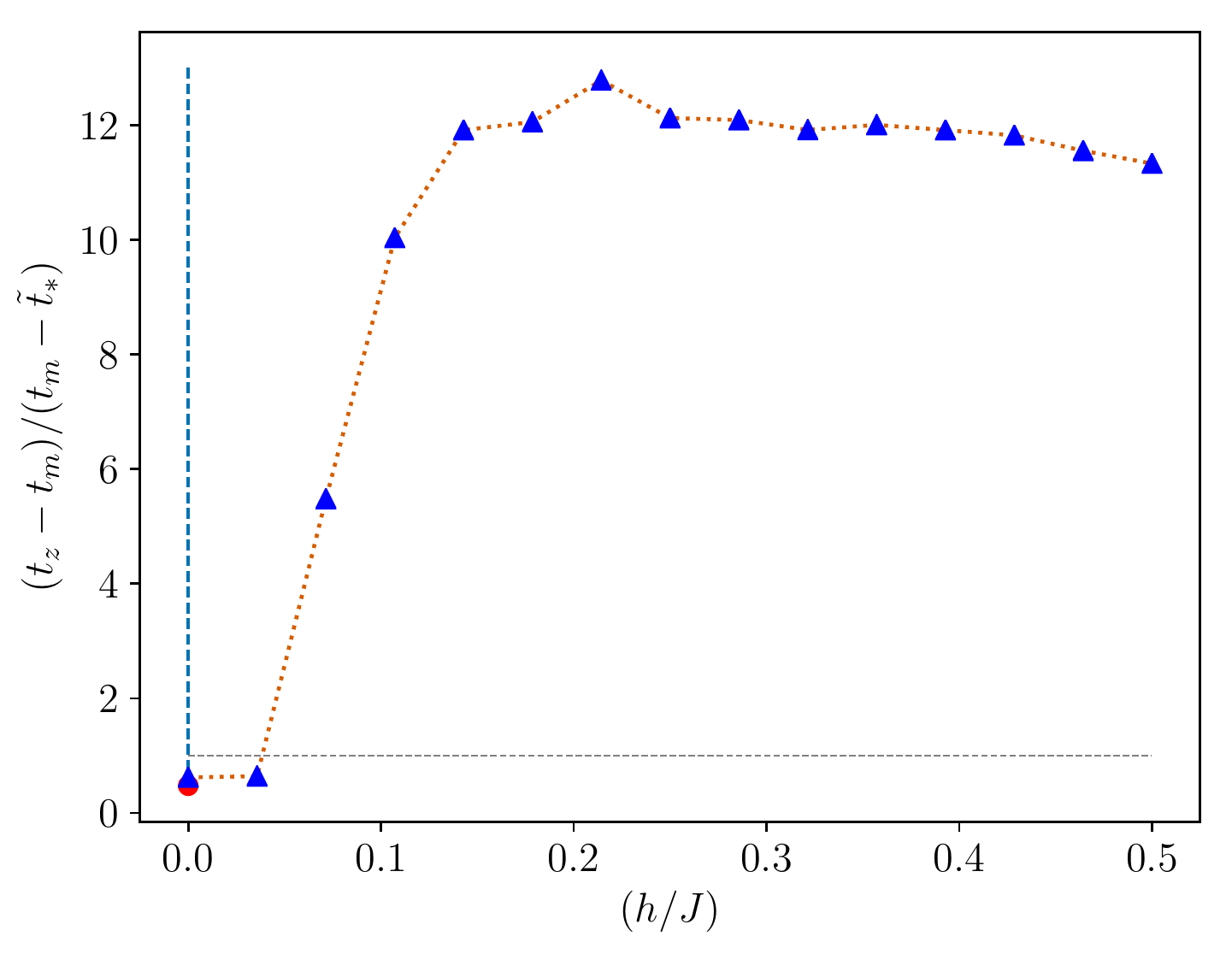} 
        }
        \subfloat[Finite temperature \label{fig:ratio_inds_T_comp}]{
            \includegraphics[width=0.45\linewidth]{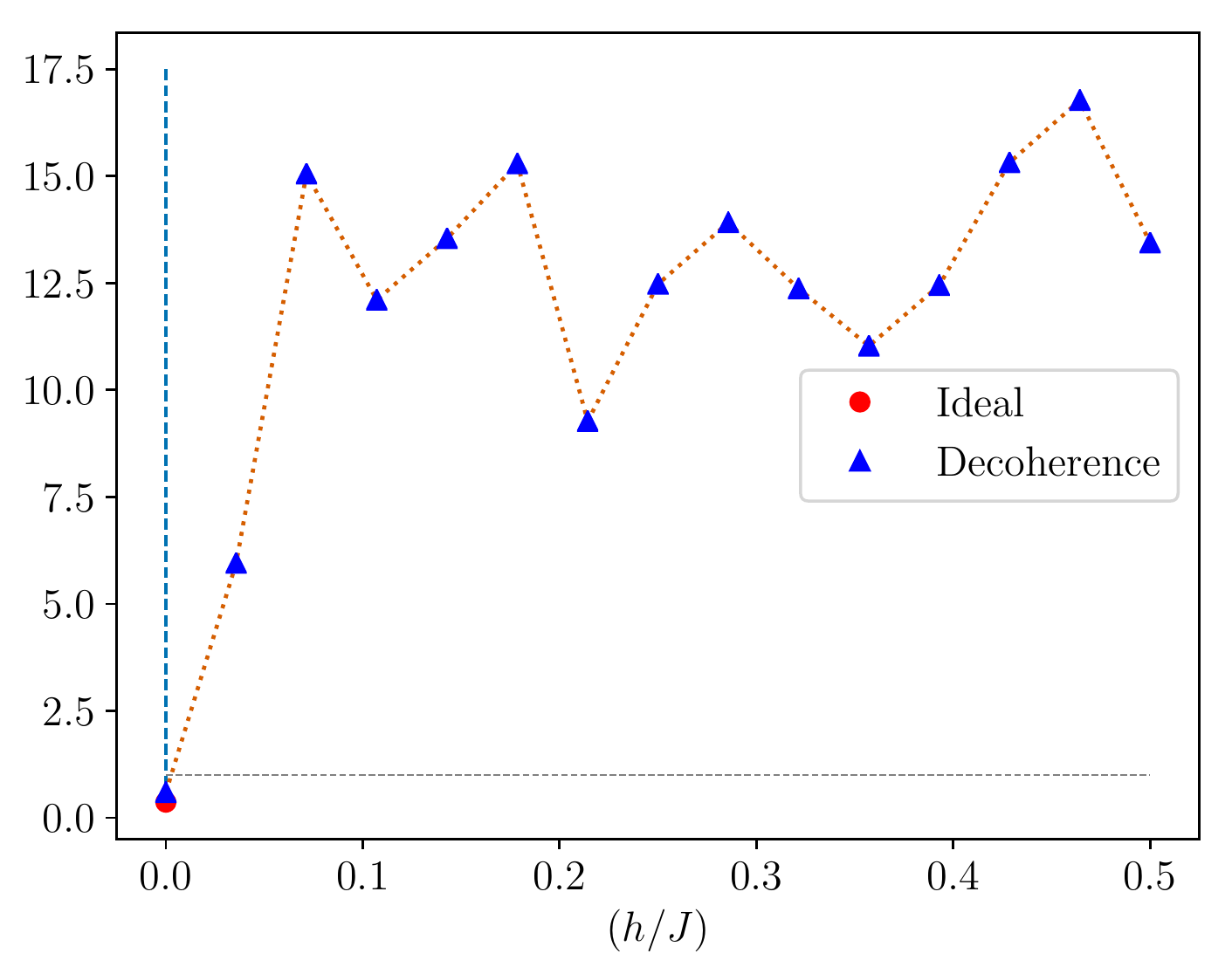}
        }
    \caption{Behavior of the ratio $(t_z - t_m)/(t_m - \tilde{t}_*)$ as a function of $h/J$ for the (a) infinite-temperature and (b) finite-temperature initial Gibbs states, $\mathrm{e}^{-H/T}/Z$, with and without decoherence. Recall that $t_z$ is longer than the total simulation time for all the decoherence-free nonintegrable cases $(h/J \ne 0)$. Therefore, the only ideal (decoherence-free)
    case we report is integrable ($h/J=0$).
}\label{fig:ratio_inds_comp}
\end{figure*}

We illustrate our conjecture using different values of $h/J$ in Fig.~\ref{fig:ratio_inds_comp}, where we plot the ratio
$(t_z - t_m)/(t_m - \tilde{t}_*)$ as a function of $h/J$. We interpret the quantity $t_z - t_m$ as a measure of how long it takes for 
some quantum information in the system to recollect, whereas $t_m - \tilde{t}_*$ indicates the time to achieve maximal quasiprobability 
nonclassicality.
Therefore, their ratio is a measure of how asymmetrical the first peak in the total nonclassicality $\tilde{N}(t)$ is.
There is a noticeable difference between the integrable and nonintegrable cases: For the decoherence-free case, 
there is a discontinuous transition where the recollection time $t_z-t_m$ becomes longer than the simulation time for
$h/J > 0$. Adding decoherence softens this transition. The ratio remains of order 1 for a wider range of small $h/J$,
in accordance with the expectation of integrability, before a sharp but smooth transition to a ratio that is over an order
of magnitude larger, in accordance with the expectation of non-integrability.

\begin{figure*}[htbp]
        \subfloat[Infinite temperature \label{fig:prev_zero_inds_inf_comp}]{
           \includegraphics[width=0.45\linewidth]{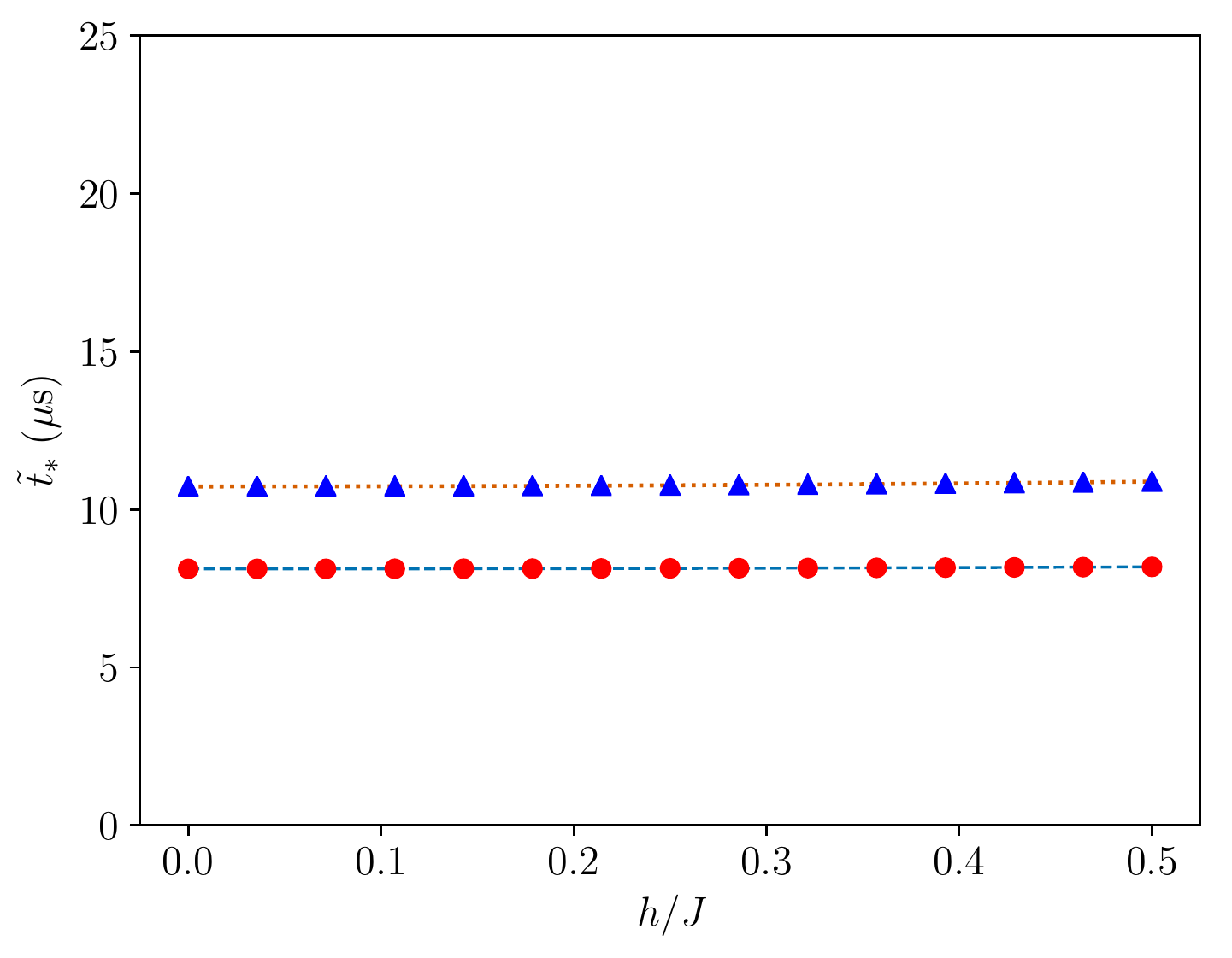} 
        }
        \subfloat[Finite temperature \label{fig:prev_zero_inds_T_comp}]{
            \includegraphics[width=0.45\linewidth]{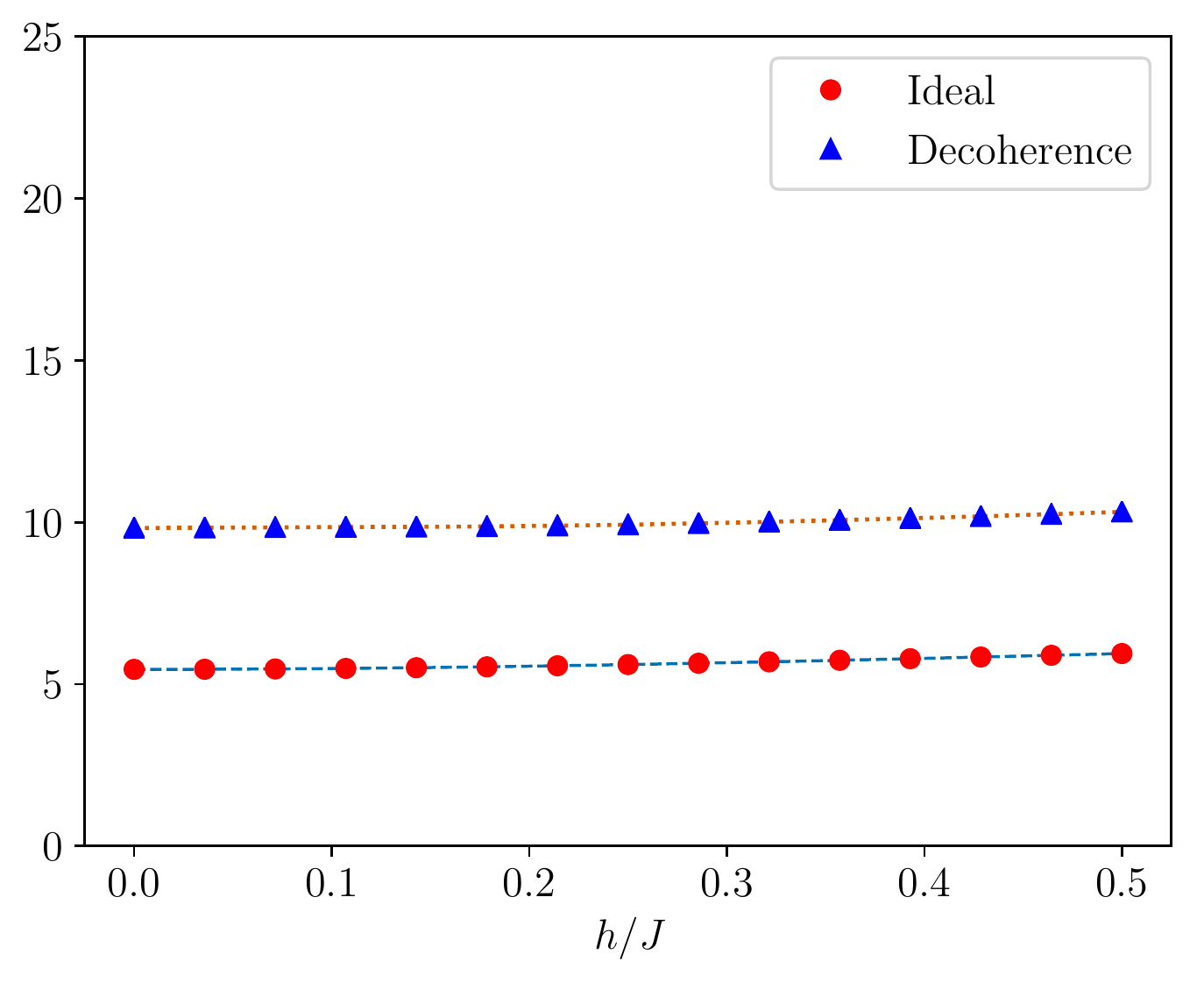}
        }
    \caption{Behavior of $\tilde{t}_*$, the point at which $\tilde{N}(t)$ first deviates from zero, as a function of $h/J$
    for the (a) infinite-temperature and (b) finite-temperature initial Gibbs states, with and without decoherence.
    Simulation parameters are similar to those in previous examples.}\label{fig:prev_zero_inds_comp}
\end{figure*}

To gain further insights into the behavior of the ratio $(t_z - t_m)/(t_m - \tilde{t}_*)$, we additionally
study the behavior of each of the time scales $\tilde{t}_*$, $t_m$, and $t_z$ separately.
Figure \ref{fig:prev_zero_inds_comp} shows that $\tilde{t}_*$, the point at which $\tilde{N}(t)$ first deviates
from zero, is hardly affected by changes in $h/J$. To compute $\tilde{t}_*$ in our simulations, we
detected the first deviation from a bound set by the square of the time step $\Delta t$ used in our numerical
simulations. However, as Fig.~\ref{fig:prev_zero_inds_comp} shows, the onset of nonclassicality is delayed by 
decoherence. This is to be expected, since, in the presence of decoherence, it is more difficult for the
system to build the coherence responsible for nonclassical behavior. Furthermore,
for a fixed value of $h/J$, systems with an infinite-temperature initial Gibbs state tend to have values of
$\tilde{t}_*$ larger than their counterparts with a finite-temperature initial Gibbs state. The infinite-temperature
state is initially diagonal in the eigenblocks of $W$ and $V$ and therefore requires more time to build quantum coherences 
than the finite-temperature state.

\begin{figure*}[htbp]
        \subfloat[Infinite temperature \label{fig:first_max_inds_inf_comp}]{
           \includegraphics[width=0.45\linewidth]{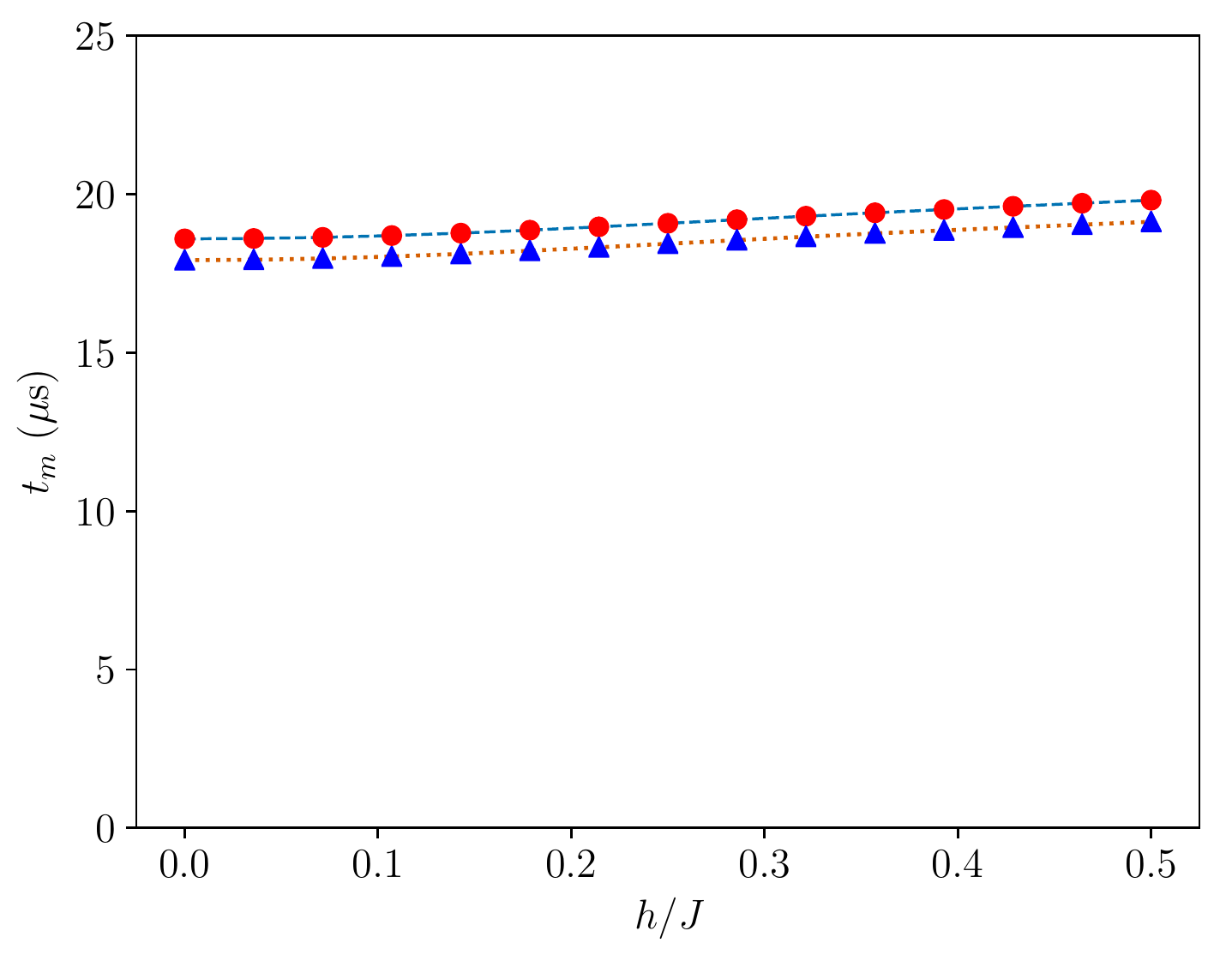} 
        }
        \subfloat[Finite temperature \label{fig:first_max_inds_T_comp}]{
            \includegraphics[width=0.45\linewidth]{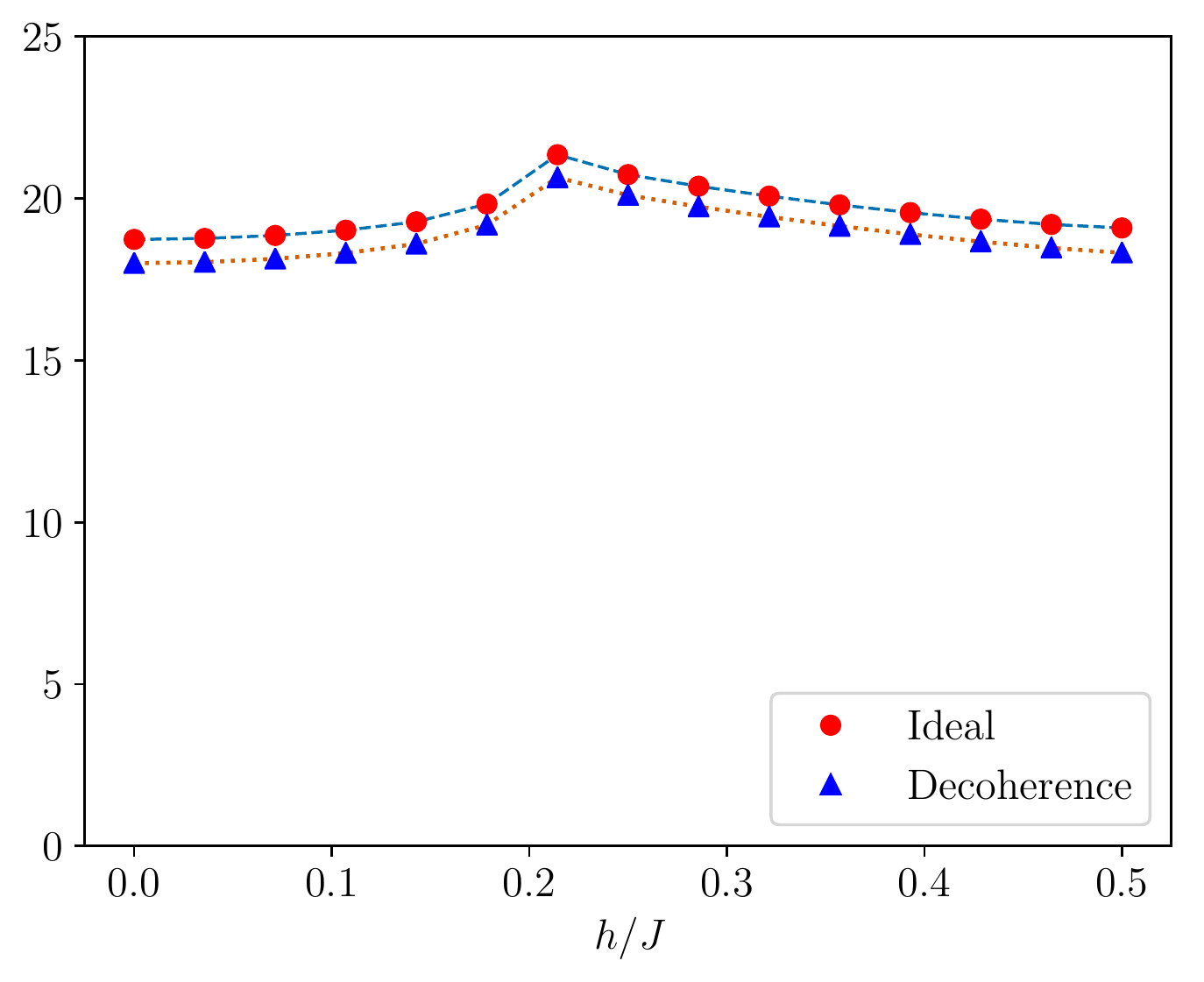}
        }
    \caption{Behavior of $t_m$, the point at which $\tilde{N}(t)$ reaches its first local maximum, as a function of $h/J$ for the
    (a) infinite-temperature and (b) finite-temperature initial Gibbs states, with and without decoherence.}\label{fig:first_max_inds_comp}
\end{figure*}

Next, in Fig.~\ref{fig:first_max_inds_comp}, we present the behavior of the point in time, $t_m$, at which the first maximum
in $\tilde{N}(t)$ as a function of $h/J$ occurs. Decoherence decreases the time required to reach
the first maximum. This is reasonable, since decoherence overall dampens the nonclassicality and therefore reduces the value of
the maximum. Hence, it becomes easier for the system to reach the smaller value of $\tilde{N}(t)$ in a shorter amount
of time. $t_m$ depends on the system dynamics and the initial state. We can appreciate an interesting difference in the plots 
corresponding to different choices of initial state. Whereas the curves corresponding to the infinite-temperature initial Gibbs states 
show a monotonic behavior, the ones for the finite-temperature initial Gibbs states briefly rise and then fall. Further understanding of 
this particular behavior is left for future research.

\begin{figure*}[htbp]
        \subfloat[Infinite temperature \label{fig:next_zero_inds_inf_comp}]{
           \includegraphics[width=0.45\linewidth]{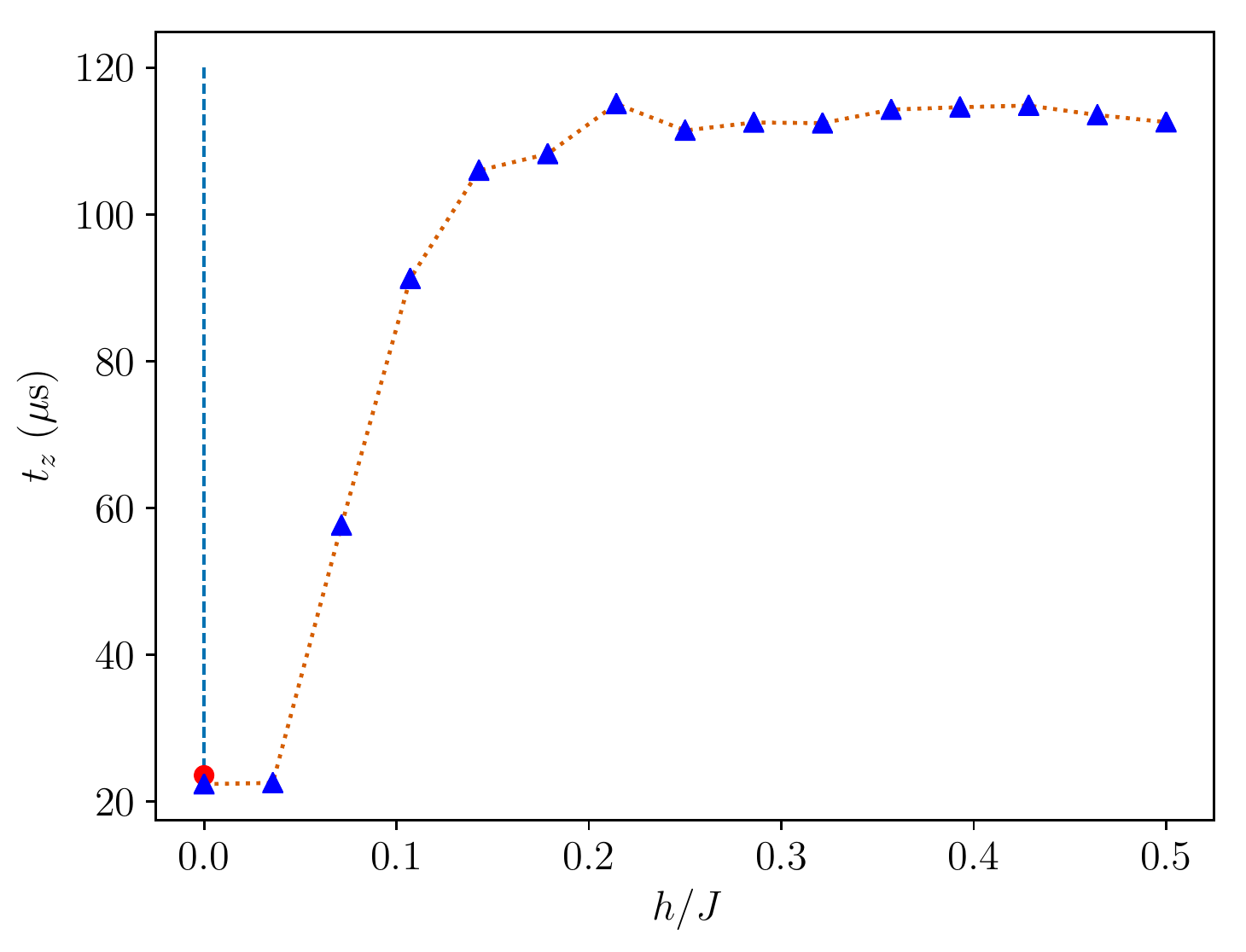} 
        }
        \subfloat[Finite temperature \label{fig:next_zero_inds_T_comp}]{
            \includegraphics[width=0.45\linewidth]{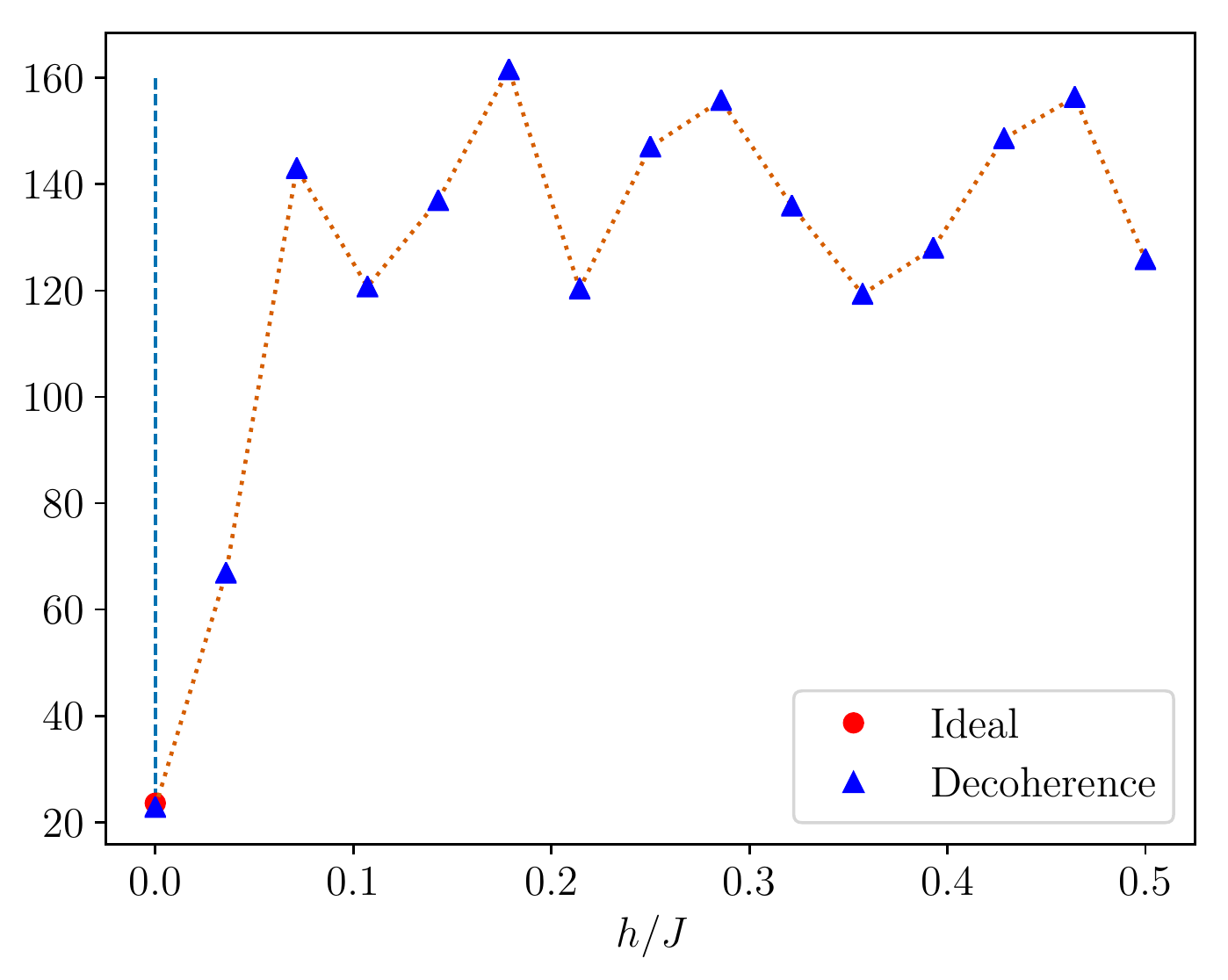}
        }
    \caption{Behavior of $t_z$, the point at which $\tilde{N}(t)$ reaches a subsequent zero after $t_m$, as a function of $h/J$ for the
    (a) infinite-temperature and (b) finite-temperature initial Gibbs states with and without decoherence.
    For all the nonintegrable cases $(h/J \ne 0)$ without decoherence, $t_z$ was longer than the total simulation time.}\label{fig:next_zero_inds_comp}
\end{figure*}

Finally, in Fig.~\ref{fig:next_zero_inds_comp}, we observe how the time $t_z$ to the the subsequent zero after the first maximum changes 
with $h/J$. As expected, without decoherence, only $h/J=0$ reaches zero again in a time shorter than the total simulation time, i.e., the 
maximum value of $t$ for which we calculated $F(t)$, $\tilde{p_t}$, and $\tilde{N}(t)$. However, this changes with the addition of 
decoherence. With it, $t_*$, the time to reach zero again, is shorter than the total simulation time for all cases. However, $t_*$ is 
significantly longer for the nonintegrable cases, regardless of the choice of initial state.

\bibliography{../references/Reference_Database}